\documentclass[aps,prd,10pt,showpacs,amsmath,twocolumn,floatfix,amssymb,nofootinbib,longbibliography]{revtex4-2}

\usepackage{graphicx}
\usepackage{comment}
\usepackage[usenames]{color}
\usepackage{bm}
\usepackage{ifpdf}
\usepackage{floatrow}
\usepackage{makecell}
\usepackage{caption}
\usepackage{stackrel}
\usepackage{subcaption}
\usepackage{enumitem}

\usepackage[normalem]{ulem}
\usepackage[dvipsnames]{xcolor}
\usepackage[utf8]{inputenc}
\usepackage{hyperref}
\hypersetup{
  pdfnewwindow=true,      
  colorlinks=true,        
  linkcolor=PineGreen,    
  citecolor=PineGreen,    
  filecolor=PineGreen,    
  urlcolor=PineGreen      
}
\newcommand{\be}{\begin{eqnarray}}
\newcommand{\ee}{\end{eqnarray}}

\usepackage{parskip}
\usepackage{anyfontsize}

\begin{document}

\title{Bloch oscillation with a diatomic tight-binding model on quantum computers}

\author{Peng~Guo}
\email{peng.guo@dsu.edu}

\affiliation{College of Arts and Sciences,  Dakota State University, Madison, SD 57042, USA}
\affiliation{Kavli Institute for Theoretical Physics, University of California, Santa Barbara, CA 93106, USA}

\author{Jaime~Park}
\email{jaime.s.park@vanderbilt.edu}

\affiliation{ Department of Computer Science, Vanderbilt University, Nashville, TN 37235, USA}

\author{Frank~X.~Lee}
\email{fxlee@gwu.edu}
\affiliation{Department of Physics, George Washington University, Washington, DC 20052, USA}

\date{\today}

\begin{abstract}
We aim to explore a more efficient way to simulate few-body dynamics on quantum computers. Instead of mapping the second quantization of  the system Hamiltonian to qubit Pauli gates representation via the Jordan-Wigner transform,  we propose to use  the few-body Hamiltonian matrix under the statevector basis representation which is more economical on the required number of quantum registers. For a single-particle excitation state on a one-dimensional chain, $\Gamma$ qubits can simulate $N=2^\Gamma$ number of sites, in comparison to $N$ qubits for $N$ sites via the Jordan-Wigner approach. A two-band diatomic tight-binding model is used to demonstrate the effectiveness of the statevector basis representation. Both one-particle and two-particle quantum circuits are constructed and some numerical tests on IBM hardware are presented.
\end{abstract}

\maketitle

\section{Introduction}\label{sec:intro}

 Quantum computing offers the prospect to simulate complex quantum systems that are  inaccessible by classical approaches. It also  has the potential to  overcome intrinsic limitations facing classical simulations, such as critical slowing down  \cite{SCHAEFER201193} in lattice Quantum Chromodynamics (QCD), no access to real-time dynamics,  sign problem in strongly interacting fermions with non-zero density~\cite{PhysRevB.41.9301,deForcrand:2009zkb}, and signal-to-noise ratio \cite{lepage1989analysis,DRISCHLER2021103888} in multi-baryon correlation functions in lattice QCD calculations. The general strategy of simulating a physical system on a quantum computer is to map the system Hamiltonian onto an effective  model defined by quantum qubits.  Then the time evolution of the physical system can be programmed in steps through unitary gate operations, see e.g. the review in \cite{https://doi.org/10.1002/qute.201900052}. Take a simple system with fermion interactions, the Hubbard model~\cite{doi:10.1098/rspa.1963.0204}, as an example. 
 Simulation of such a system on a universal digital quantum computer typically involves,
\begin{itemize}
\item map the second quantization representation of system Hamiltonian onto qubit Pauli gates representation via the Jordan-Wigner transform  \cite{PhysRevA.64.022319,PhysRevA.65.042323,JordanWigner};
\item translate unitary time evolution operator of the system into a set of quantum gates, and evolve the system in small time steps through the Trotterization approximation \cite{TrotterH.F.1959Otpo,Hatano:2005gh};
 \item prepare many-body initial state, see e.g. Ref.~\cite{PhysRevA.64.022319}, and then evolve  the state by applying quantum circuit of unitary time evolution operator in the previous step;
 \item measure  physical observables, such as expectation values and correlation functions, e.g. Refs.~\cite{PhysRevA.64.022319,PhysRevA.65.042323,https://doi.org/10.1002/qute.201900052};
 \item assess possible errors in the results due to noise in the quantum hardware, a step called {\em error mitigation} and {\em error correction}.
\end{itemize}

Mapping system Hamiltonian via Jordan-Wigner transform  \cite{PhysRevA.64.022319,PhysRevA.65.042323,JordanWigner} onto qubit Pauli gates representation may be suitable for simulating {\em many-body} systems, but it is not the most efficient way for {\em few-body} dynamics.  For instance, a $N$-site 1D tight-binding spinless two-fermion Hamiltonian can be mapped onto $N$ quantum registers via Jordan-Wigner transform, hence the size of Hamiltonian matrix is $2^N \times 2^N$. 
As will be detailed in this work, for a single-particle excitation, the same $N$-site Hamiltonian under {\em statevector basis representation} can be mapped onto $\Gamma = \log_2 N$ quantum registers, hence single-particle Hamiltonian matrix has the size of $N\times N$. For two particles excitation, two sets of $\Gamma$ quantum registers are required, hence the size of Hamiltonian matrix grows to $N^2 \times N^2$. Similarly,  $q$ sets of $\Gamma$ quantum registers are required for $q$-particle systems.   Admittedly, simulating few-body systems under statevector basis representation becomes inefficient at the point when the size of Hamiltonian $N^q \times N^q$ outgrows the matrix size $2^N \times 2^N$ of Hamiltonian under Jordan-Wigner representation. For few-body dynamics, statevector basis representation has clear advantages in terms of efficiency on the usage of qubits.

We choose to work with a simple but still interesting quantum system: the two-band   diatomic tight-binding  1D chain model. This model exhibits the well-known  phenomenon of {\em Bloch oscillation}~\cite{Bloch1929,doi:10.1098/rspa.1934.0116}:  electrons placed on a crystal lattice and subject to an external electric field are not accelerated across the lattice, but undergo oscillatory motion in a localized region of the lattice.  The localization behavior is also referred to as the Wannier-Stark localization \cite{RevModPhys.34.645}.  Bloch oscillation and Wannier-Stark localization reduce electron transport on the crystal lattice drastically. The dispersion relation of electrons in the model forms ladders that are referred to as Wannier-Stark ladders \cite{Hartmann_2004,callaway1974quantum}. In the continuum theory,  Wannier-Stark ladder states become Wannier-Stark resonance states, see e.g. the review in Ref.~\cite{GLUCK2002103}. Bloch oscillation has been experimentally observed  in semiconductor superlattices \cite{PhysRevB.46.7252}, cold atom in an optical lattice \cite{PhysRevLett.76.4508}, and photonic crystal \cite{PhysRevLett.83.4752,PhysRevLett.83.4756}. The observation of single-particle excitation  Bloch oscillation and Wannier-Stark localization on a superconducting quantum processor was also reported recently \cite{SongPRX2024,GuoNature2021}.

We will show that for the diatomic tight-binding model, the few-body Hamiltonian can be mapped to qubit basis efficiently. Only two sets of $\Gamma = \log_2 N$  quantum registers are required to simulate two interacting fermions for a $N$-site 1D system.  The hopping terms, electric field term, and interaction term in the system Hamiltonian can be mapped onto simple quantum circuits.  Their exact expressions for unitary time evolution can be obtained.  Only a few qubits and simple quantum circuits are required to simulate a modest-sized system.

The paper is organized as follows. After introduction in Sec.~\ref{sec:intro},  the two-band diatomic tight-binding model and its few-body Hamiltonians under  statevector basis representation  are discussed    in Sec.~\ref{sec:tightbindingmodel}.  Their quantum circuit realizations for single-particle and two-particle states are presented in Sec.~\ref{sec:quantumcircuits}.   Numerical results,  discussions and summary are given in Sec.~\ref{sec:numerics} and  Sec.~\ref{sec:summary} respectively. Some technical details are given in two appendices.

\section{Two-band  tight-binding model}\label{sec:tightbindingmodel}

\begin{figure}[b]
\includegraphics[width=0.95\textwidth]{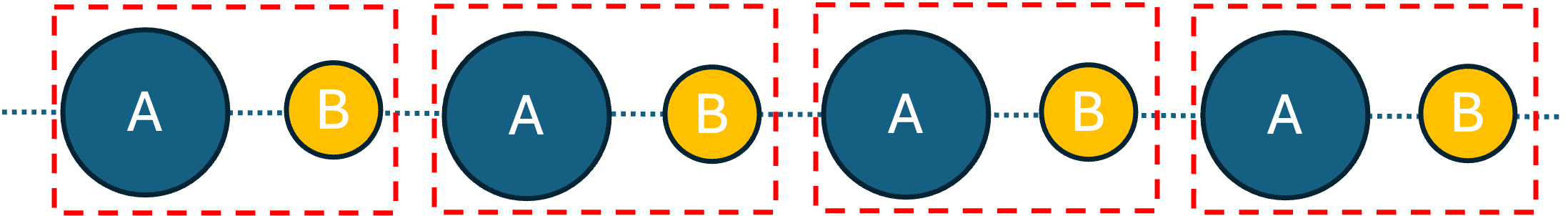}   
\caption{An illustration of the diatomic tight-binding model. It is a one-dimensional, open-ended chain that can be truncated to a finite system of $N$ sites with periodic boundary conditions for numerical studies.}
\label{fig:model}
\end{figure}

We consider a simple diatomic tight-binding model on a one-dimensional, open-ended chain, as shown in Fig.~\ref{fig:model}.  Two types of atoms are placed in each unit cell with type-A on even sites and type-B on odd sites.  An electron can hop between type-A and type-B atoms within a single cell with hopping constant $\triangle_a/4$,  or between two neighboring cells with hopping constant $\triangle_b/4$.   The simplest realization of such an $(AB) -(AB)$ chain-like system in nature with different hoppings between intra- and extra-cell  would be the polyyne type of periodic $C-C\equiv C-C$ structure with alternating single and triple bonds in-between carton chains.   In the presence of a time-dependent  linear  electric  potential (which corresponds to a spatially-uniform electric field along the chain),  the second quantization representation of tight-binding model Hamiltonian has three contributions,
\begin{equation}
 \hat{H}  (t) =  \hat{H}_0 + \hat{H}_{E} (t) + \hat{H}_V      .  \label{HtwobandTB}
\end{equation}
The first term,   $\hat{H}_0   $, is free electron Hamiltonian in the absence of electric field and particle interaction,
\begin{align}
& \hat{H}_0   
=    - \frac{\triangle_a}{4}  \sum_{n = -\infty}^{\infty}  \sum_{s = \uparrow, \downarrow  }   \left [ a^\dag_{2 n, s} b_{2 n+1, s}  + b^\dag_{2 n+1, s} a_{2 n, s} \right ] \nonumber \\
&  -   \frac{\triangle_b}{4}   \sum_{n = -\infty}^{\infty} \sum_{s = \uparrow, \downarrow  }   \left [ b^\dag_{2 n+1, s} a_{2 n+2, s} + a^\dag_{2 n +2, s} b_{2 n+1, s} \right ]  ,  \label{H0twobandTB}
\end{align}
where $a^\dag_{2n, s}$ and $b^\dag_{2n+1, s}$    refer to creation operators of  electrons with   polarization $s$ (spin-up or spin-down).  The second term, $\hat{H}_{E} (t)$, describes the  linear electric potential   of strength $F(t)$,
\begin{equation}
 \hat{H}_E  (t)
=     F (t)   \sum_{l = -\infty}^{\infty}  \sum_{s = \uparrow, \downarrow  }   l \,   c^\dag_{l, s} c_{l, s}  ,  \label{HEtwobandTB}
\end{equation}
where 
\begin{equation}
c_{l ,s}^\dag  = \begin{cases} a_{l ,s}^\dag , &\mbox{if} \ \  l  = 2 n , \\ b_{l ,s }^\dag , & \mbox{if} \ \  l = 2 n +1.   \end{cases}
\end{equation}  
A time-dependent external electric field with both dc and ac components,
\begin{equation}
F(t) = F + F_{ac} \cos ( \omega t),
\end{equation}
is adopted to simulate the dynamic localization effect that is driven by the ac component, see e.g. Ref.~\cite{Holthaus01081996}.
The last term, $ \hat{H}_V  $, represents the  contact interaction between electrons with opposite spin polarizations,
\begin{equation}
 \hat{H}_V  
=    V  \sum_{l  = -\infty}^{\infty}    c^\dag_{l , \uparrow} c_{l , \uparrow}    c^\dag_{l , \downarrow} c_{l , \downarrow}  ,  \label{HUtwobandTB}
\end{equation}
where $V$ stands for the potential strength of the contact interaction.

\subsection{Single-particle state in non-interacting cases}
For non-interacting cases ($V=0$), the polarization of an electron is conserved, hence the spin index $s$ will be suppressed in this subsection, and the electron is treated as a spinless fermion. For numerical simulations, the infinite 1D chain is truncated into a finite-size system with $N$ sites and periodic boundary conditions.

\subsubsection{Jordan-Wigner representation}

In this representation, the non-interacting tight-binding model Hamiltonian operator in Eq.(\ref{HtwobandTB})    truncated to $N$ sites can  be mapped onto  $N$-qubit quantum registers by applying the Jordan-Wigner transformation, see e.g.  Refs.~\cite{PhysRevA.64.022319,PhysRevA.65.042323,JordanWigner}. For example, for a spinless fermion, creation  operators are mapped to
\begin{equation}
c_l^\dag =I_{N-1} \otimes \cdots  \otimes I_{l+1} \otimes \sigma^\dag_l \otimes Z_{l -1} \otimes \cdots  \otimes Z_{0} ,   
\end{equation}
where $\sigma_l^{\pm} = \frac{X_l \mp i Y_l}{2}$, $(X,Y,Z)$ denote Pauli-($X,Y,Z$) gates respectively, $I$ is the identity gate, and  the   subscript  $l$ is used to  label $l$-th quantum register.  
 The Hamiltonian operator  is turned into that of a many-body spin model, 
\begin{align}
& \hat{H}_{jw}  (t)
=    - \frac{\triangle_a}{4}  \sum_{n = 0}^{\frac{N}{2}-1}    \left [ \sigma_{2n+1}^+ \sigma_{2n}^- + \sigma_{2n+1}^- \sigma_{2n}^+ \right ] \nonumber \\
&  -   \frac{\triangle_b}{4}   \sum_{n = 0}^{\frac{N}{2}-1}   \left [ \sigma_{2n+2}^+ \sigma_{2n+1}^- + \sigma_{2n+2}^- \sigma_{2n+1}^+\right ]  \nonumber \\
&  +    F (t)   \sum_{l  = 0}^{N-1}   l \,   \sigma_{l}^+ \sigma_{l}^-   .  \label{HqsTB}
\end{align}
The $ \hat{H}_{jw}  (t)$ now can be carried out in terms of Pauli-gate qubit operations on quantum computers.  With $N$ qubits for $N$ sites 1D chain,  the matrix size of $ \hat{H}_{jw}  (t)$  is $2^N \times 2^N$ in Jordan-Wigner representation.

The single-particle excitation state is defined by, 
\begin{equation}
| \Psi (t) \rangle = \mathcal{T} \left [ e^{- i \int_0^t d t'  \hat{H}_{qs}  (t')} \right ]  | \Psi (0) \rangle = \sum_{l  = 0}^{N-1} \psi (l  , t)  c_l^\dag | 0 \rangle  ,
\end{equation}
where $ \mathcal{T} $ denotes the time-ordered operator.
The amplitude squared at site-$l$ can be measured by
\begin{equation}
 | \psi (l  , t)  |^2 = \langle  \Psi (t ) | \sigma_l^+ \sigma_l^- | \Psi (t) \rangle   .
\end{equation}

\subsubsection{Statevector basis representation}
In this representation, the individual single-particle excitation at site-$l$ is defined by
\begin{equation}
|  l \rangle  = c_l^\dag | 0 \rangle = \begin{bmatrix} 0_0 & \cdots & 0_{l -1}   &  1_l & 0_{l +1}  & \cdots & 0_{N-1} \end{bmatrix}^T,  \label{singleparticlestatebasis}
\end{equation} 
thus the overall single-particle state is given by
 \begin{equation}
| \Psi (t) \rangle  = \sum_{l  = 0}^{N-1} \psi (l  , t)  | l \rangle = \begin{bmatrix}  \psi (0 , t) \\  \psi (1 , t) \\ \vdots   \end{bmatrix}  ,
\end{equation}
where $ |\psi (l  , t) |^2$ describes the probability of finding a particle at site-$l$ at time $t$. For $N$-site finite size single-particle system, only $\Gamma = \log_2 N$ quantum registers are required.  As the number of qubits is increased, the size of finite system increase exponentially.
See Ref.~\cite{Guo:2025vgk} for an application of the statevector basis representation in a scattering problem.

In terms of statevector basis, the Hamiltonian matrix for the single-particle excitation is thus given by,
\begin{equation}
\hat{H}_{sv} (t ) \equiv \hat{H}_a + \hat{H}_b +  \hat{H}_E(t),
\label{Hftonebody}
\end{equation}
 where  
 \begin{align}
  \hat{H}_{a} &=  - \frac{\Delta_a}{4} \sum_{n=0}^{\frac{N}{2}-1}  \left ( | 2 n \rangle \langle  2 n +1 | +  | 2 n +1 \rangle \langle  2 n  |  \right ) ,   \nonumber \\
  \hat{H}_{b}  &=   - \frac{\Delta_b}{4} \sum_{n=0}^{\frac{N}{2}-1}  \left ( | 2 n +1 \rangle \langle  2 n +2 | +  | 2 n +2 \rangle \langle  2 n +1 |  \right ) , 
  \label{eq:Hab}
  \end{align}
  and
  \begin{equation}
 \hat{H}_{E} (t )    =  F (t)   \sum_{l  = 0}^{N-1}   l \,  | l \rangle \langle l |  .
  \label{eq:HE}
 \end{equation}
With $\Gamma $ qubits for $N=2^\Gamma$ sites,  the matrix size of $ \hat{H}_{sv}(t)$  is $N\times N$ in statevector basis representation.

The Schr\"odinger equation for single-particle excitation state,
\begin{equation}
i \frac{\partial}{\partial t}  | \Psi (t) \rangle = \hat{H}_{sv}  (t) | \Psi (t) \rangle,
\end{equation}
leads to  the following coupled differential equations,
\begin{align}
& \left [ i \frac{\partial }{ \partial t}  - F(t) 2 n \right ]  \psi(2n, t)  \nonumber \\
 & \qquad = - \frac{\Delta_b}{4} \psi(2n-1, t)  - \frac{\Delta_a}{4} \psi(2n+1, t)      , \nonumber \\
& \left [ i \frac{\partial}{\partial t} - F(t) (2 n+1) \right ] \psi(2n+1, t)  \nonumber \\
&\qquad = - \frac{\Delta_a }{4} \psi(2n, t)  - \frac{\Delta_b}{4} \psi(2n+2, t)  ,    \label{diffeqtwoband}
\end{align}
where $ n \in [0, \frac{N}{2} -1] $ for a finite system of $N$-site 1D chain, along with periodic boundary condition $\psi(0,t) = \psi(N,t)$. Eq.(\ref{diffeqtwoband}) can be solved numerically.  To facilitate quantum circuit design, we introduce a notation to project out  the single-particle wavefunction amplitude squared at site-$l$ by,
\begin{equation}
 | \psi (l  , t)  |^2 = \langle  \Psi (t ) | \hat{P} (l) | \Psi (t) \rangle   ,
\end{equation}
where $ \hat{P} (l) $ is the projection operator of a diagonal matrix with only a single no-zero element at site-$l$:
\begin{equation}
 \hat{P} (l) = | l \rangle \langle l | = \begin{bmatrix} 
 0_0 &  \cdots & 0 &   \cdots     &   0  \\
\cdots &  \cdots &  \cdots &     \cdots  & \cdots   \\
 0&   \cdots &  1_l & \cdots           &0 \\
\cdots &  \cdots &  \cdots &     \cdots  & \cdots \\
  0  & \cdots & 0&   \cdots  &  0_{N-1}
 \end{bmatrix} .
 \label{eq:p1}
\end{equation}

Some physics properties of the diatomic non-interacting tight-binding model are given in Appendix \ref{sec:basicproperties}. Our objective here is to simulate the system on a quantum computer.

\subsection{Statevector basis representation of two-particle state in interacting cases}

With contact  potential between two electrons interacting with opposite spins ($V \neq 0$),  we consider   two-electron system in singlet state of total spin ($S=0, S_z =0$), defined on finite sites  by
\begin{equation}
| \Psi (t) \rangle = \sum_{l_1, l_2 =   0 }^{N-1} \psi(l_1, l_2, t) | l_1, l_2 \rangle, \label{twoparticlestatedef}
\end{equation}
where
\begin{equation}
| l_1, l_2 \rangle = \frac{1}{\sqrt{2}} \left (  c^\dag_{l_1, \uparrow}  c^\dag_{l_2, \downarrow} -  c^\dag_{l_1, \downarrow}  c^\dag_{l_2, \uparrow}   \right ) | 0 \rangle .
\end{equation}
Using the state in Eq.(\ref{twoparticlestatedef}) and the Schr\"odinger equation, 
the kinetic term of the non-interacting Hamiltonian for two-particle state  can be mapped to  
\begin{equation}
\hat{H}_{K} (t)= \hat{H}^{(1)}_{sv} (t)  \otimes   I^{(2)}  +    I^{(1)}   \otimes  \hat{H}^{(2)}_{sv} (t) , \label{HKtwobody}
\end{equation}
where superscripts in $ \hat{H}^{(i)}_{sv} (t)$ and $  I^{(i)}$ with $i=1,2$ are used to label  the single-particle Hamiltonian and operator for $i$-th   particle, single-particle Hamiltonian operator $ \hat{H}_{sv} (t)$ is defined in Eq.(\ref{Hftonebody}) and $  I^{(i)}$ denotes the unit operator for $i$-th particle: 
\begin{equation}
  I^{(i)}  =   \sum_{l_i = 0 }^{ N-1 } | l_i \rangle  \langle l_i |  .
\end{equation}
The contact potential term of Hamiltonian is mapped to
\begin{equation}
\hat{H}_V = V \sum_{l_1, l_2 =   0 }^{N-1}  \delta_{l_1, l_2} | l_1, l_2  \rangle  \langle l_1, l_2  |  . \label{HVtwobody}
\end{equation}

The two-particle basis $| l_1, l_2 \rangle$ now can be mapped to the tensor product of two single-particle statevector basis:
\begin{equation}
| l_1, l_2 \rangle  = | l_1 \rangle \otimes | l_2 \rangle ,
\end{equation}
where single-particle statevector basis is defined in terms of column matrix in Eq.(\ref{singleparticlestatebasis}). Hence, two-particle state  Hamiltonian operators $\hat{H}_{K} (t)$ and $\hat{H}_V$ in Eq.(\ref{HKtwobody}) and Eq.(\ref{HVtwobody}) respectively are now given by matrix representation in terms of two-particle statevector basis. The matrix representation of $\hat{H}_{K} (t)$ is given by the tensor product of $\hat{H}_{sv} (t) $ matrix in Eq.(\ref{Hftmatrixrep}) and unit matrix, and the   $\hat{H}_V $   is given by a diagonal matrix of size $N^2 \times N^2$,
\begin{equation}
\hat{H}_{V}   = \begin{bmatrix} 
\begin{pmatrix}
1 & 0 & \cdots \\
0 & 0 & \cdots \\
\cdots & \cdots & \cdots
\end{pmatrix}_{N\times N} & 0_{N\times N} & \cdots \\
0_{N\times N} &  \begin{pmatrix}
0 & 0 & \cdots \\
0 & 1 & \cdots \\
\cdots & \cdots & \cdots
\end{pmatrix}_{N\times N}  & \cdots \\
\cdots & \cdots & \cdots
 \end{bmatrix}  .
\end{equation}

The two-particle wavefunction amplitude squared at site-$(l_1, l_2)$ can be projected out by
\begin{equation}
 | \psi (l_1, l_2  , t)  |^2 = \langle  \Psi (t ) | \hat{P} (l_1, l_2) | \Psi (t) \rangle   ,
\end{equation}
where
\begin{equation}
 \hat{P} (l_1, l_2) =  \hat{P} (l_1) \otimes    \hat{P} (l_2)= \left  ( | l_1 \rangle \langle l_1 | \right ) \otimes   \left  ( | l_2 \rangle \langle l_2 | \right ),
\end{equation}
is the two-particle projection operator built from the tensor product of single-particle operators in Eq.\eqref{eq:p1}.

\section{Quantum circuits of two-band tight-binding model in statevector basis representation}\label{sec:quantumcircuits}

Having laid out the physics details of the Bloch oscillation in the two-band tight-binding model in the statevector basis representation, we now seek to build quantum circuits for its realization on quantum computers. 
We first use single-particle state to illustrate the basic elements, then proceed to two-particle states.

\subsection{Quantum circuits of single-particle system}

\begin{figure}[h]
\frame{\includegraphics[width=.49\textwidth]{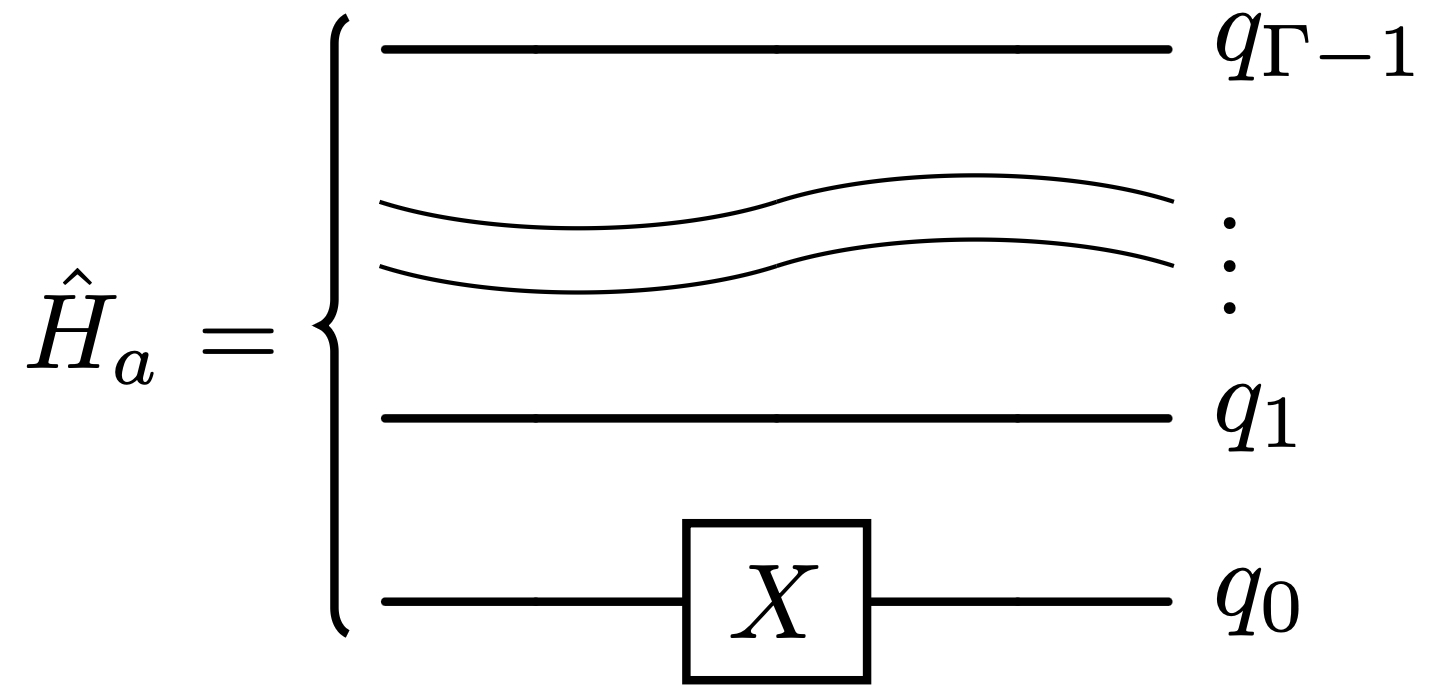}}
\caption{Quantum circuit of $\hat{H}_a$ in Eq.\eqref{eq:Hab}.}
\label{FIGHa}
\end{figure}

\begin{figure}[h]
\frame{\includegraphics[width=.99\textwidth]{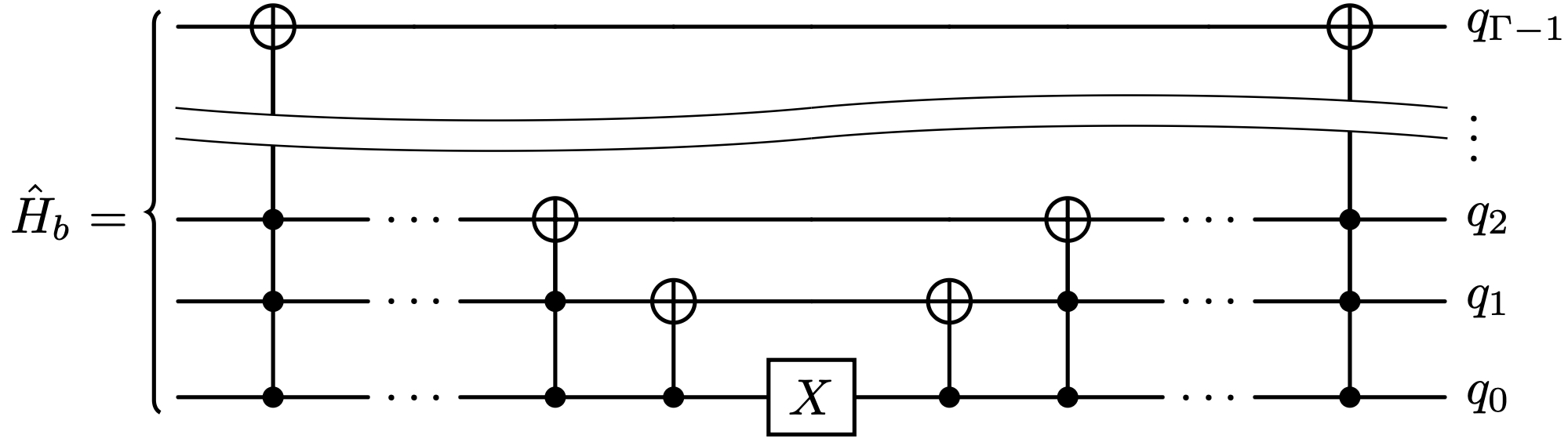}}
\caption{Quantum circuit of $\hat{H}_b$ in Eq.\eqref{eq:Hab}.}
\label{FIGHb}
\end{figure}

\begin{figure}[h]
\frame{\includegraphics[width=.90\textwidth]{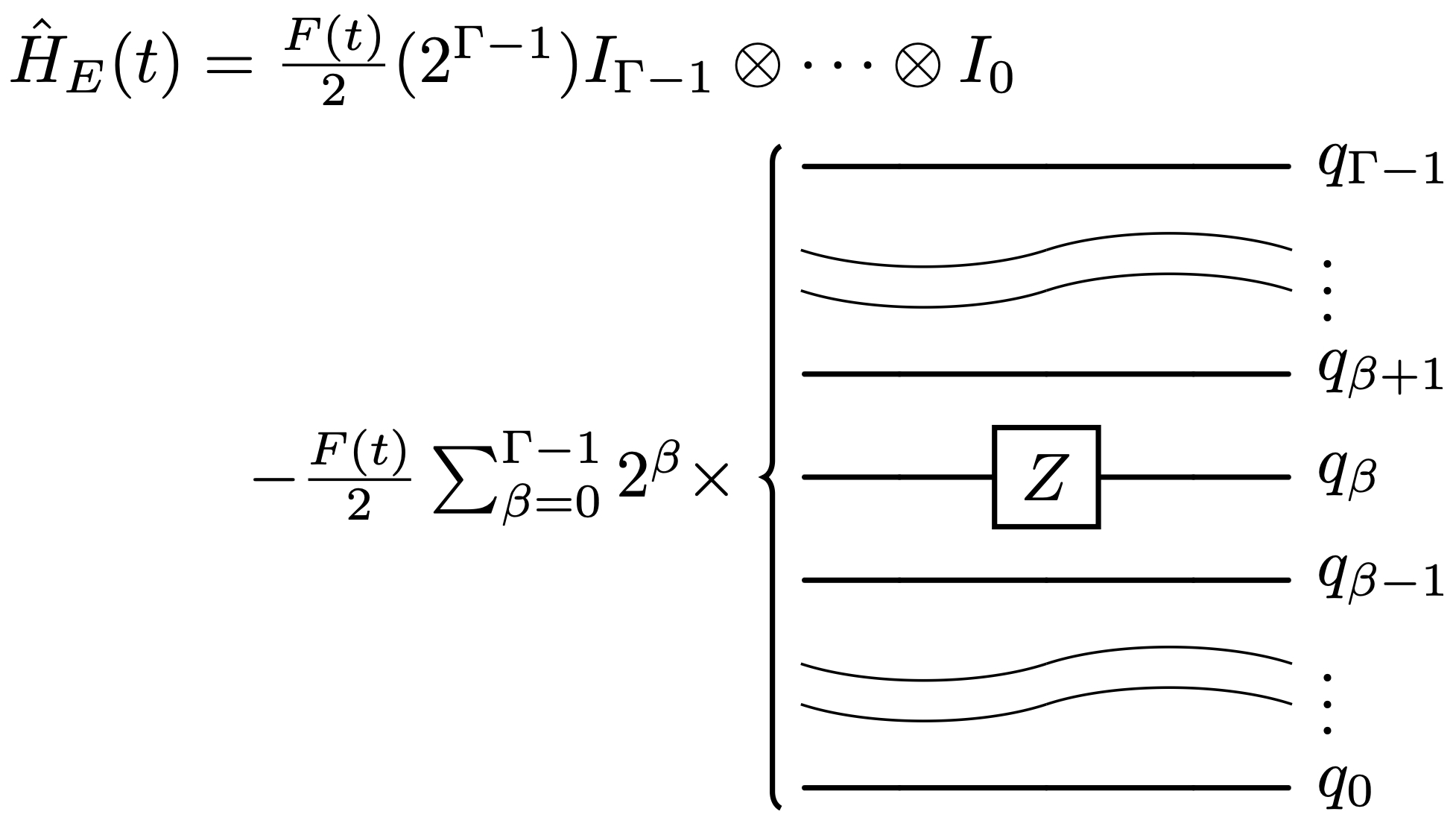}}
\caption{Quantum circuit of $\hat{H}_{E} (t)$ in Eq.\eqref{eq:HE}. The second term represents the sum of all possible single $Z$-gate insertions with weight factors that depend on the position. For example, with one $Z$-gate insertion at  $\beta$-th qubit,  the weight factor  is $\frac{F(t)}{2} 2^\beta$.}
\label{FIGHE}
\end{figure}

After some tedious algebra, we constructed quantum circuits of the single-particle Hamiltonian matrices in Eq.(\ref{Hftonebody}) to Eq.\eqref{eq:HE}, or equivalently its matrix form Eq.(\ref{Hftmatrixrep}). For a $N$-site system, it can be mapped onto $\Gamma$-qubit quantum registers, where $2^\Gamma = N$.  Specifically, $\hat{H}_{a}$ and $\hat{H}_{b}$ are given in terms of $X$-gate and multi-controlled-$X$ gates in Fig.~\ref{FIGHa} and Fig.~\ref{FIGHb}, respectively.
And  $\hat{H}_{E}(t)$ is given in terms of linear superposition of   $Z$-gate circuits in Fig.~\ref{FIGHE}.

\begin{figure}[h]
\frame{\includegraphics[width=.79\textwidth]{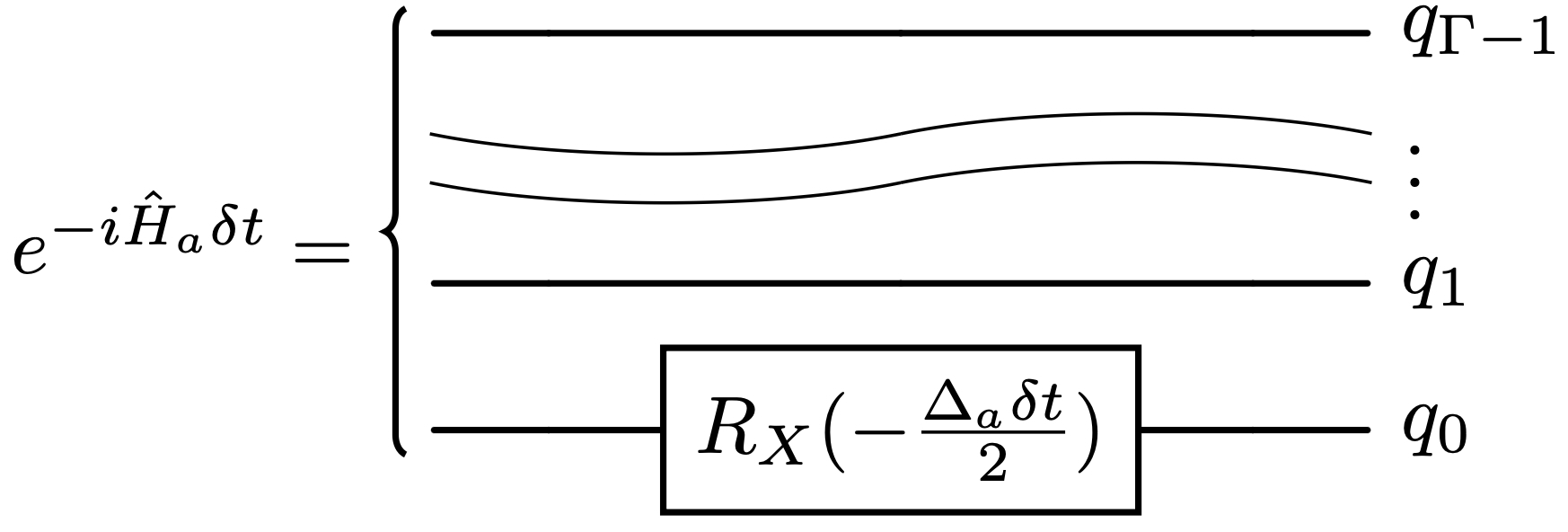}}
\caption{Quantum circuit of $e^{-i \hat{H}_a \delta t }$ in Eq.\eqref{eq:eHa}.}
\label{FIGexpHa}
\end{figure}

\begin{figure}[h]
\frame{\includegraphics[width=.99\textwidth]{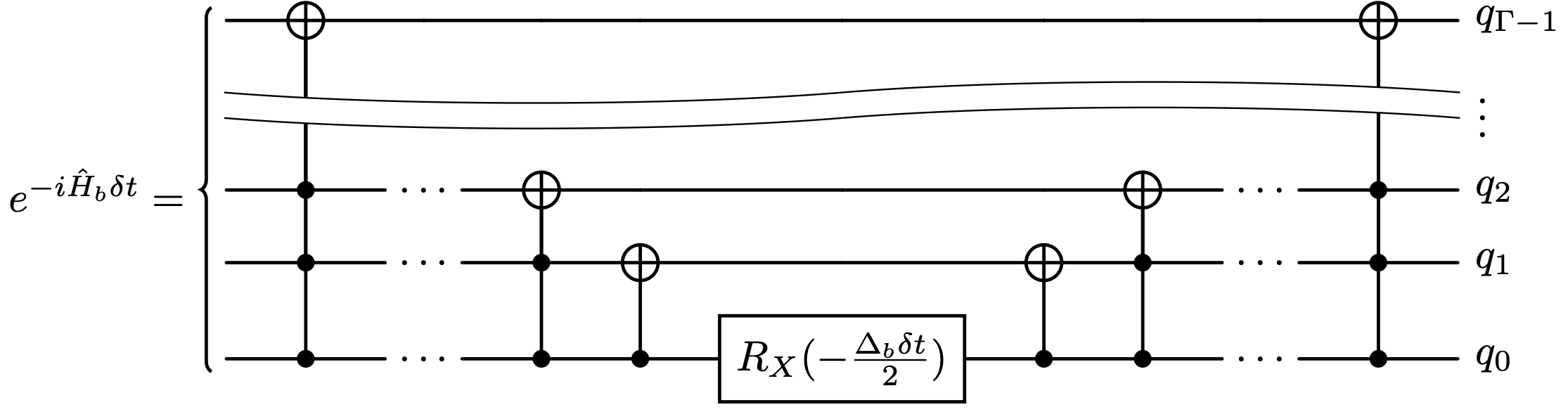}}
\caption{Quantum circuit of $e^{-i \hat{H}_b \delta t }$ in Eq.\eqref{eq:eHb}.}
\label{FIGexpHb}
\end{figure}

\begin{figure}[h]
\frame{\includegraphics[width=.95\textwidth]{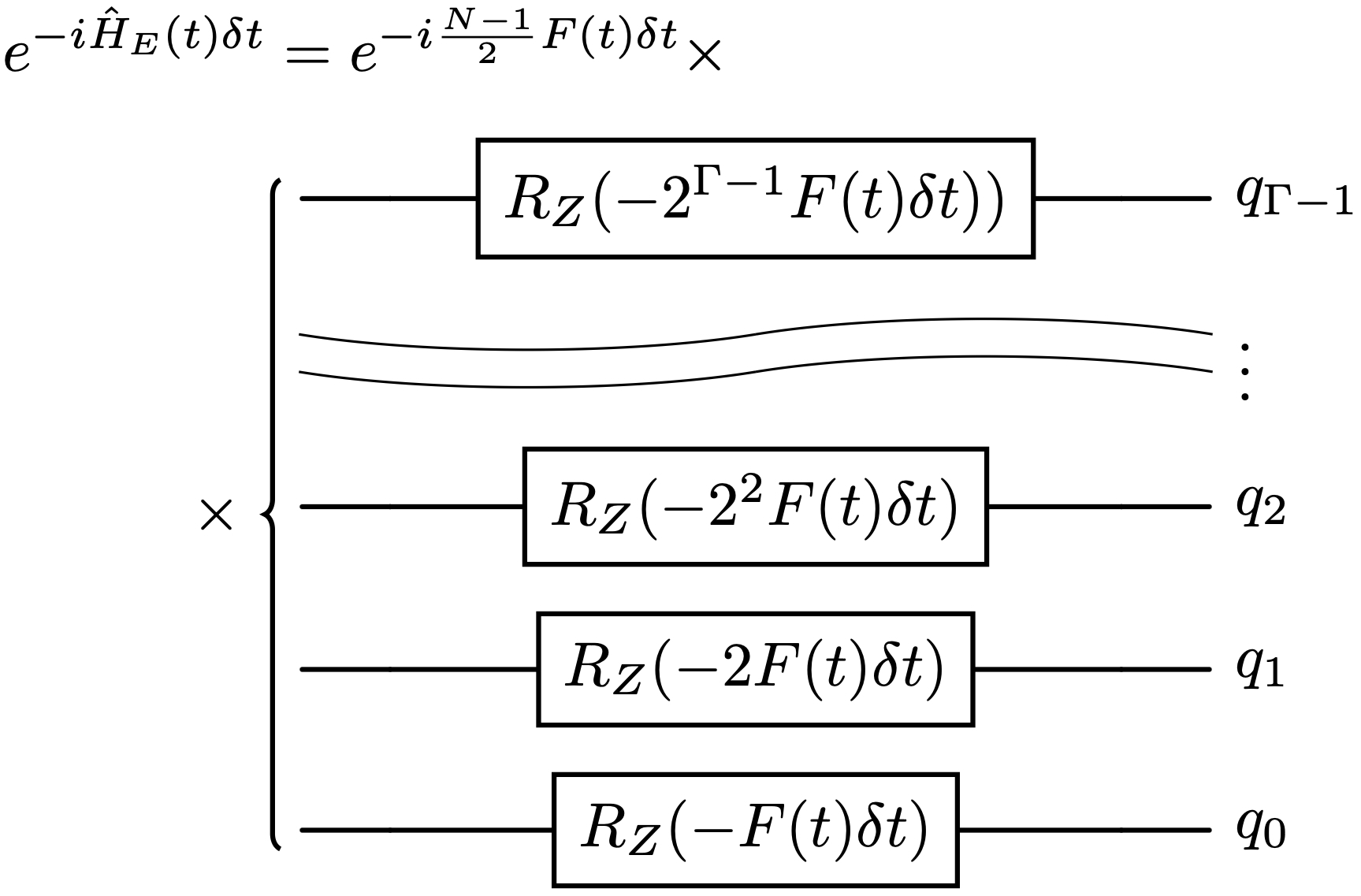}}
\caption{Quantum circuit of $e^{-i \hat{H}_E (t) \delta t }$ in Eq.\eqref{eq:eHE}.}
\label{FIGexpHE}
\end{figure}

The corresponding time evolution exponential of each term of $\hat{H}_{sv} (t )  $ has exact solutions,
\begin{align}
& e^{- i  \hat{H}_{a} \delta t}  \nonumber \\
& = \begin{bmatrix} 
\cos ( \frac{\Delta_a \delta t }{2} ) &  i  \sin ( \frac{\Delta_a  \delta t}{2} )  & 0 & 0 & \cdots  \\
 i  \sin ( \frac{\Delta_a  \delta t}{2} )  & \cos ( \frac{\Delta_a \delta t}{2} )  & 0 & 0 & \cdots  \\
0 & 0 &  \cos ( \frac{\Delta_a \delta t }{2} ) &  i  \sin ( \frac{\Delta_a \delta t }{2} )  & \cdots \\
0& 0 &  i  \sin ( \frac{\Delta_a \delta t }{2} )  & \cos ( \frac{\Delta_a  \delta t}{2})   & \cdots \\
\cdots & \cdots & \cdots & \cdots & \cdots
 \end{bmatrix}  ,
 \label{eq:eHa}
\end{align}
\begin{align}
& e^{- i  \hat{H}_{b} \delta t}  \nonumber \\
& = \begin{bmatrix} 
\cos ( \frac{\Delta_b \delta t }{2} ) &  0 & 0 &   \cdots & i  \sin ( \frac{\Delta_b  \delta t}{2} )  \\
0 &    \cos ( \frac{\Delta_b \delta t }{2} ) &  i  \sin ( \frac{\Delta_b \delta t }{2} )  & \cdots & 0 \\
0&   i  \sin ( \frac{\Delta_b \delta t }{2} )  & \cos ( \frac{\Delta_b  \delta t}{2})   & \cdots & 0  \\
\cdots & \cdots & \cdots & \cdots & \cdots \\
 i  \sin ( \frac{\Delta_b  \delta t}{2} )   & 0 & 0 & \cdots & \cos ( \frac{\Delta_b \delta t}{2} )  
 \end{bmatrix}  ,
  \label{eq:eHb}
\end{align}
and
\begin{align}
& e^{- i  \hat{H}_{E} (t) \delta t}  \nonumber \\
& = \begin{bmatrix} 
1 &  0 & 0 &   \cdots & 0  \\
0 &   e^{- i F(t) \delta t} &  0 & \cdots  & 0  \\
0&   0 & e^{- 2 i F(t) \delta t}   & \cdots & 0  \\
\cdots & \cdots & \cdots & \cdots & \cdots \\
 0   & 0 & 0 & \cdots & e^{- (N-1) i F(t) \delta t} 
 \end{bmatrix}  .
  \label{eq:eHE}
\end{align}
Their quantum circuits are given in Fig.~\ref{FIGexpHa}, Fig.~\ref{FIGexpHb},  and Fig.~\ref{FIGexpHE}, respectively. The exponential of the entire $\hat{H}_{sv} (t )  $  thus can be simulated on the quantum computer in small time steps via the Trotterization approximation, with the lowest order given by
\begin{equation}
U(\delta t ) = e^{- i \hat{H}_{sv} (t) \delta t  } \stackrel{ \delta t \rightarrow 0}{\approx }   e^{- i  \hat{H}_{a} \delta t}   e^{- i  \hat{H}_{b} \delta t}   e^{- i  \hat{H}_{E} (t) \delta t}  .
\label{U3}
\end{equation}

\subsection{Quantum circuits of two-particle system}

For two-particle states on the $N$-site  system,  two sets of $\Gamma = \log_2 N$ quantum registers are required to simulate two-particle Hamiltonian matrix of size $N^2 \times N^2$. The $i$-th set of $\Gamma$ qubits is used to describe the motion of $i$-th particle.  

For the non-interacting Hamiltonian $\hat{H}_K (t)$ in Eq.(\ref{HKtwobody}), the exponential of  $\hat{H}_K (t)$ is simply given by,
\begin{equation}
e^{- i  \hat{H}_K (t) \delta t } =  e^{- i   \hat{H}^{(1)}_{sv} (t)   \delta t }   \otimes   I^{(2)}  +    I^{(1)}   \otimes e^{ - i   \hat{H}^{(2)}_{sv} (t) \delta t } .
\end{equation}

\begin{figure}[h]
\frame{\includegraphics[width=.79\textwidth]{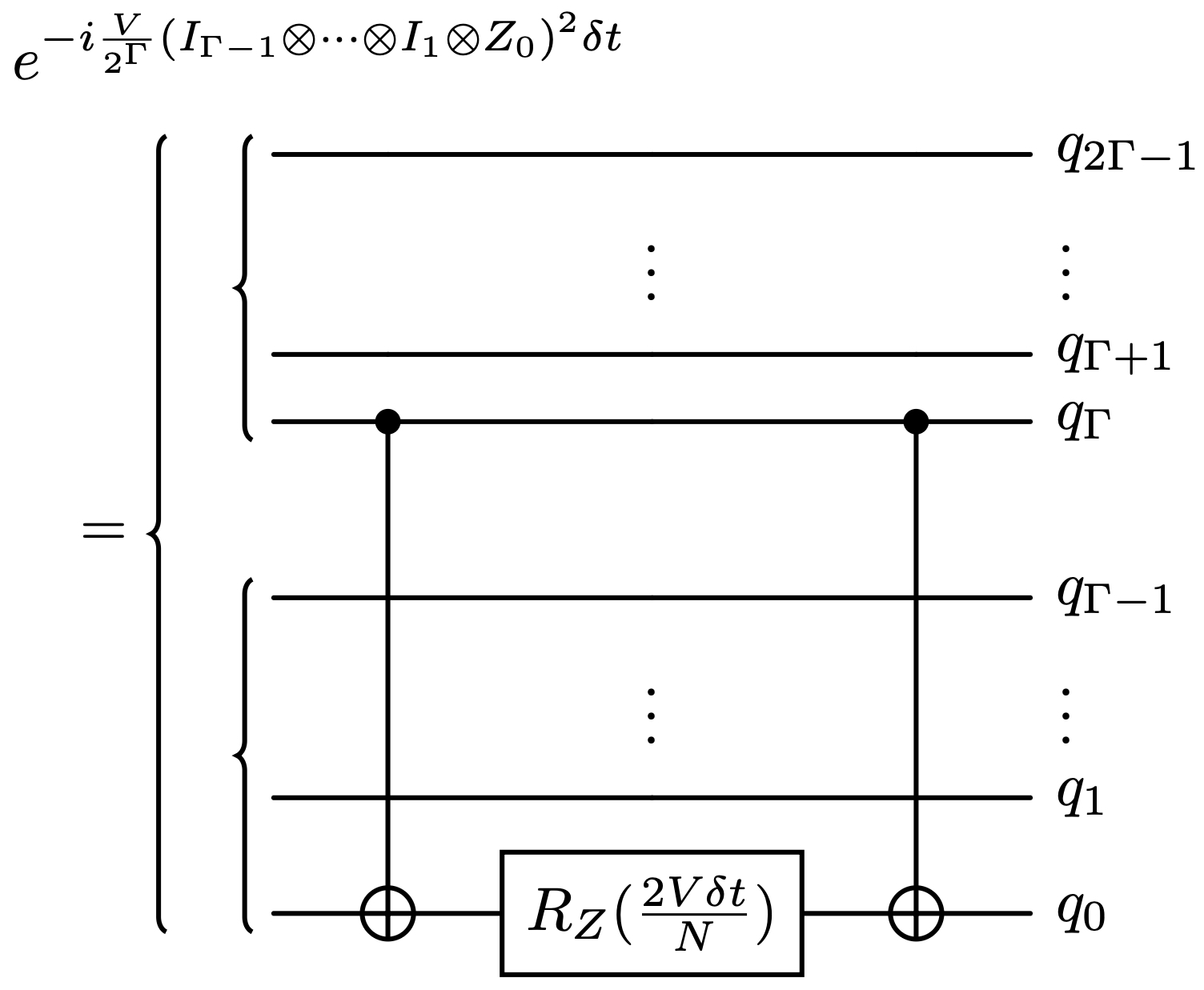}}
\caption{Quantum circuits for time evolution $e^{-i \hat{H}_V \delta t }$ of Eq.\eqref{eq:HVQC} with one $Z$-gate insertion.}
\label{FigHV1}
\end{figure}

\begin{figure}[h]
\frame{\includegraphics[width=.99\textwidth]{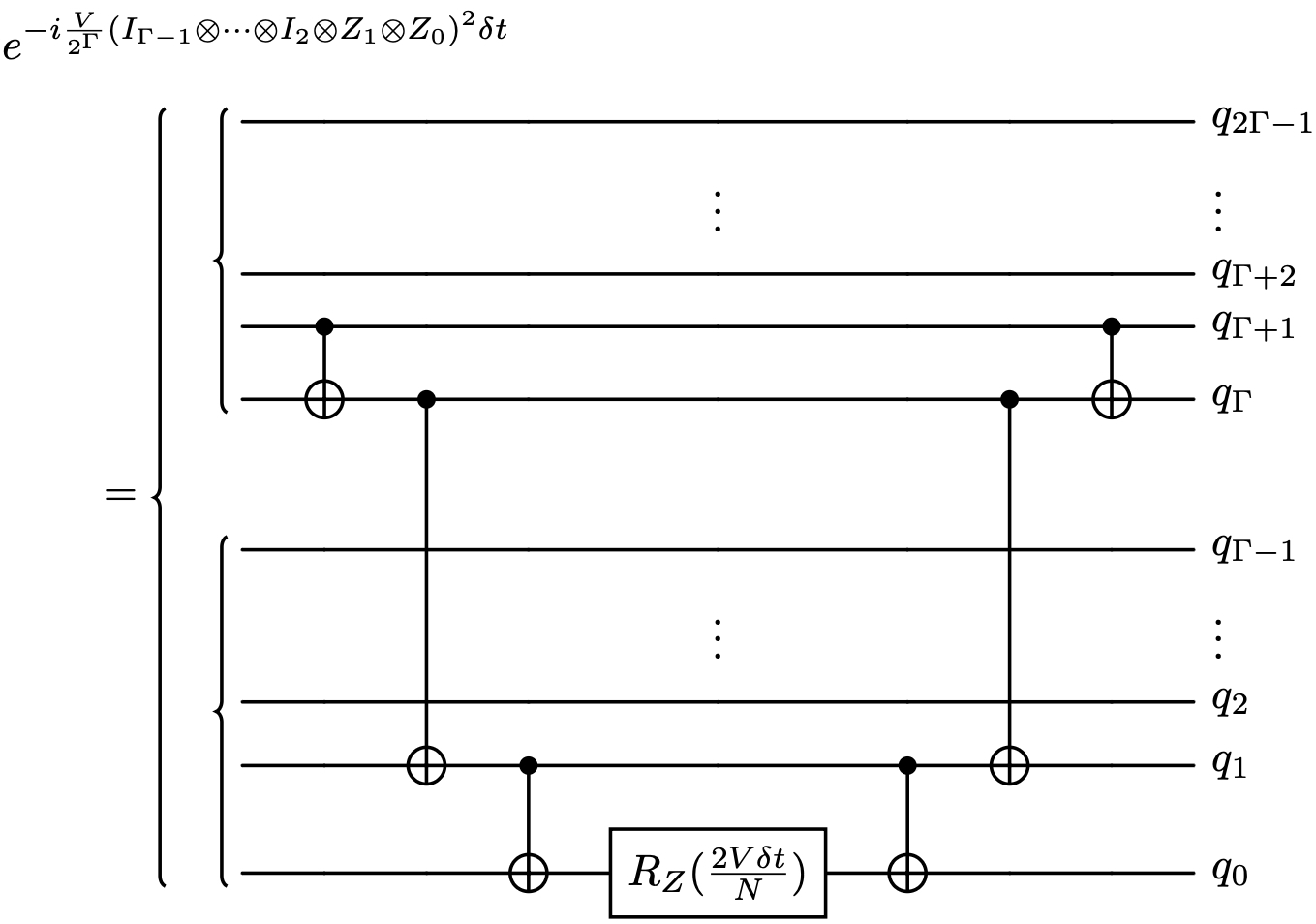}}
\caption{Quantum circuits for time evolution $e^{-i \hat{H}_V \delta t }$ of Eq.\eqref{eq:HVQC} with two $Z$-gate insertions.}
\label{FigHV2}
\end{figure}

\begin{figure}[h]
\frame{\includegraphics[width=.99\textwidth]{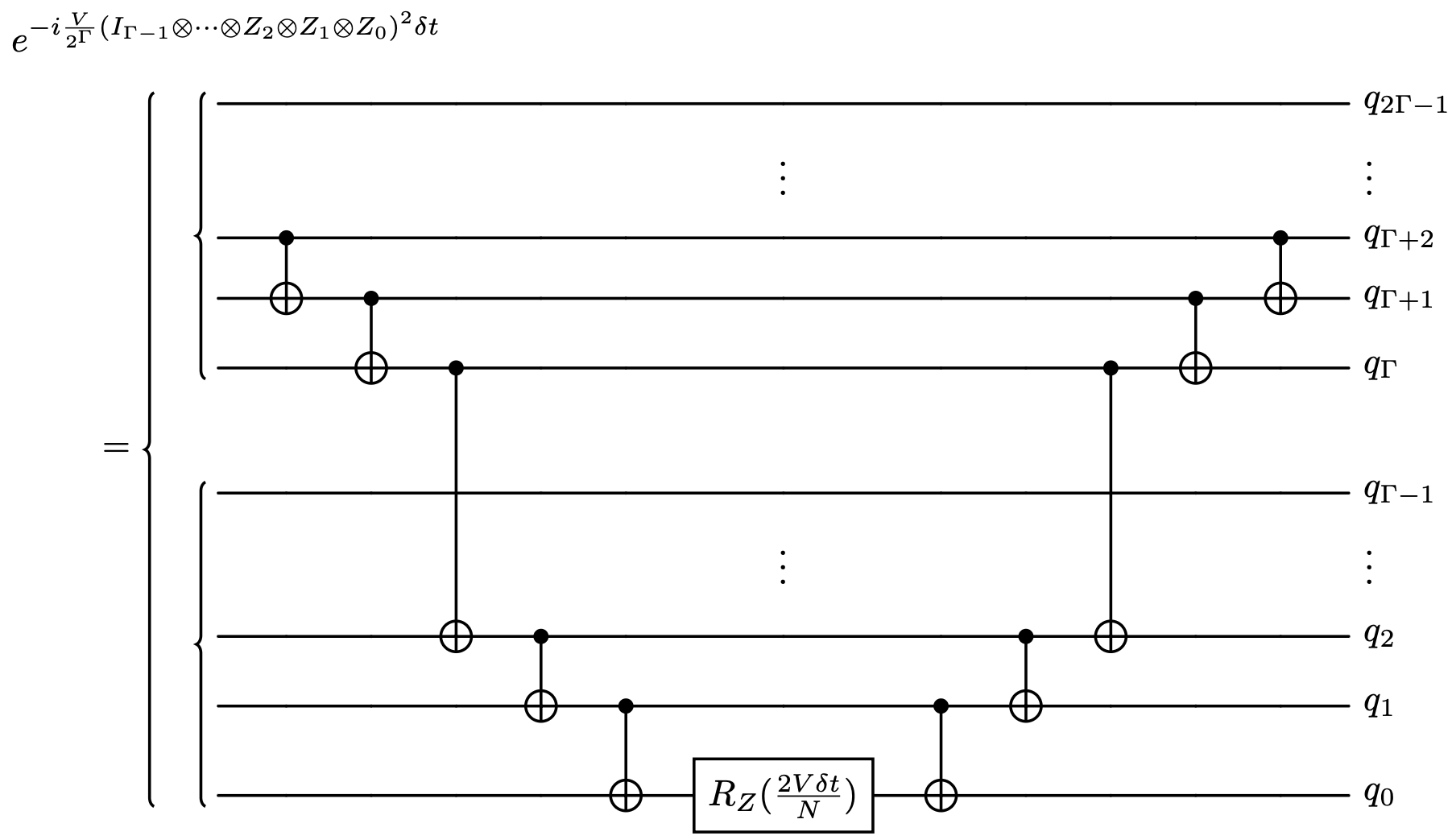}}
\caption{Quantum circuits for time evolution $e^{-i \hat{H}_V \delta t }$ of Eq.\eqref{eq:HVQC}  with three $Z$-gate insertions.}
\label{FigHV3}
\end{figure}

For the interacting Hamiltonian  $\hat{H}_V $ in Eq.(\ref{HVtwobody}), the quantum circuit of $\hat{H}_V $ in terms of $Z$-gates is given by
\begin{align}
\hat{H}_V  &= \frac{V}{2^\Gamma} \left  (I_{\Gamma-1} \otimes \cdots \otimes I_0 \right )^2  \nonumber \\
&+  \frac{V}{2^\Gamma}  \left   (I_{\Gamma-1} \otimes \cdots \otimes I_1 \otimes Z_0  \right )^2    \nonumber \\
 & + \frac{V}{2^\Gamma } \left (I_{\Gamma-1} \otimes \cdots \otimes Z_1 \otimes I_0  \right )^2     \nonumber \\
 &  + \cdots + \frac{V}{2^\Gamma } \left  (Z_{\Gamma-1} \otimes I_{\Gamma -2} \otimes \cdots \otimes  I_0 \right )^2    \nonumber \\
 &+  \frac{V}{2^\Gamma}  \left   (I_{\Gamma-1} \otimes \cdots \otimes I_2 \otimes Z_1 \otimes Z_0  \right )^2     \nonumber \\
 & + \cdots  + \frac{V}{2^\Gamma } \left  (Z_{\Gamma-1} \otimes Z_{\Gamma -2} \otimes \cdots \otimes Z_1 \otimes  Z_0 \right )^2 , \label{eq:HVQC}
\end{align}
where the square is a short-hand notation for tensor product of two sets of $\Gamma$ qubits of identical operators. The $\hat{H}_V $ is simply the linear superposition of all possible insertion of zero $Z$-gate, one $Z$-gate, \ldots, up to $\Gamma$ $Z$-gate. There are $2^\Gamma = N$ terms in  $\hat{H}_V$ given by Eq.(\ref{eq:HVQC}),   all of which  commute each other, hence 
\begin{equation}
e^{- i \hat{H}_V \delta t} = \prod_{i =1}^{N}  e^{- i \hat{H}^{(V)}_{i} \delta t},
\end{equation}
where $\hat{H}^{(V)}_{i}$ stands for each individual term in Eq.(\ref{eq:HVQC}). Therefore, quantum circuit of $e^{- i \hat{H}_V \delta t} $ is composed of $N$ simpler quantum circuits that are  given by $e^{- i \hat{H}^{(V)}_{i} \delta t}$ involving  multiple $Z$-gates. The demo plots of quantum circuits involving one, two and three $Z$-gate insertions in each set of $\Gamma$ qubits are given in Fig.~\ref{FigHV1}, Fig.~\ref{FigHV2} and Fig.~\ref{FigHV3}, respectively.

\begin{figure*}
\frame{\includegraphics[width=.95\textwidth]{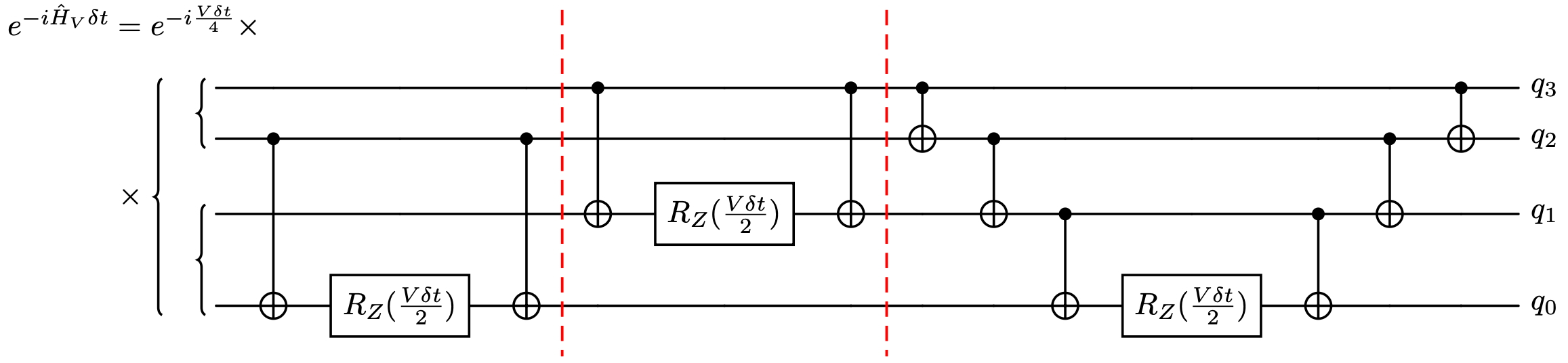}}
\caption{Quantum circuits of $e^{- i \hat{H}_V  \delta t}$ for two sets of 2-qubit.}
\label{FIGexpHV2qubit}
\end{figure*}

As a simple example, the $\hat{H}_V$ for two sets of $2$-qubit is given by
\begin{align}
\hat{H}_V  &= \frac{V}{4} \left  (I_{1}  \otimes I_0 \right )^2 +  \frac{V}{4}  \left   (  I_1 \otimes Z_0  \right )^2   + \frac{V}{4 } \left (  Z_1 \otimes I_0  \right )^2     \nonumber \\
 &   + \frac{V}{4} \left  ( Z_1 \otimes  Z_0 \right )^2 , 
\end{align}
whose corresponding quantum circuit of $e^{- i \hat{H}_V   \delta t}$ is shown in Fig.~\ref{FIGexpHV2qubit}.

\begin{figure*}
\frame{\includegraphics[width=.95\textwidth]{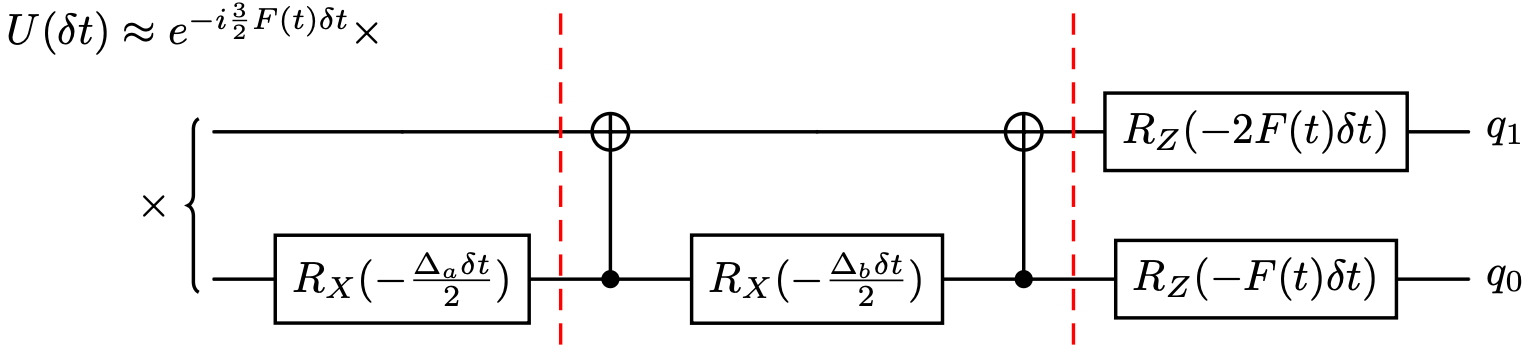}}
\caption{Quantum circuit of the lowest order $U( \delta t )$ in Eq.\eqref{U3} for  2 qubits.} \label{Fig2qubitU}
\end{figure*}

 \begin{figure*}
\frame{\includegraphics[width=.95\textwidth]{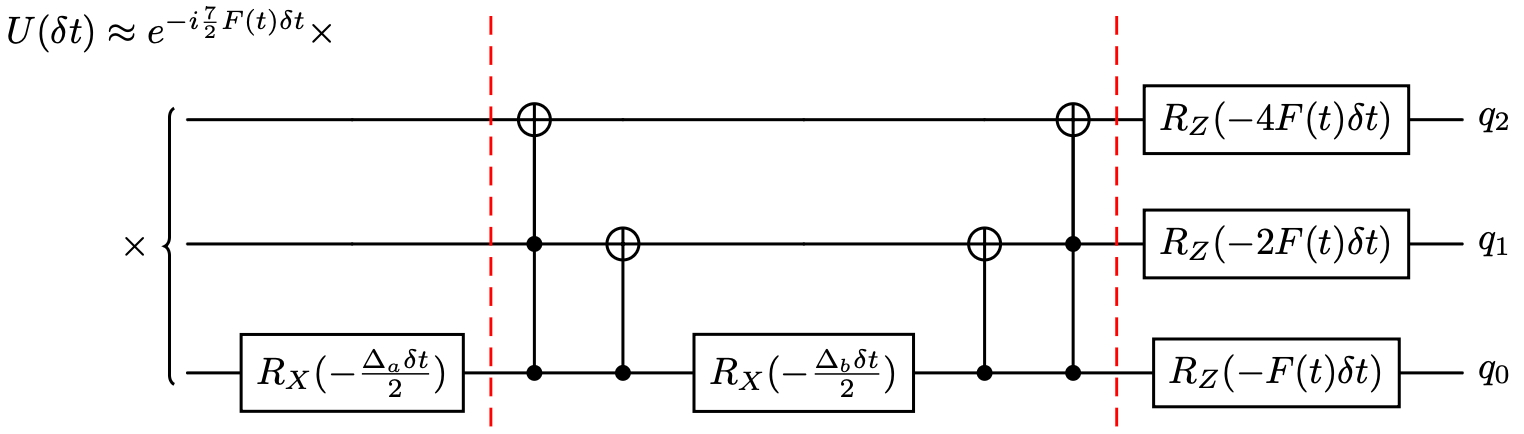}}
\caption{Quantum circuit of the lowest order $U( \delta t )$ in Eq.\eqref{U3}  for  3 qubits.} \label{Fig3qubitU}
\end{figure*}

\section{Numerics}\label{sec:numerics}

In this section, we show some numerical attempts on freely available IBM hardware (ibm\_brisbane and ibm\_sherbrooke) through cloud quantum computing access. In the following, we show the simulation results compared with exact solutions for non-interacting single-particle state cases with two and three qubits.   We also show the result for two interacting particles on a 4-site chain with four qubits (two sets of two-qubit, one set for each particle).

Two-qubit and three-qubit quantum circuits of lowest order Trotterziation approximation of Eq.\eqref{U3} 
are shown in Fig.~\ref{Fig2qubitU} and Fig.~\ref{Fig3qubitU}, respectively.  The initial state is chosen as a spike at site $l=2$, 
\begin{equation}
\psi(l, 0) = \delta_{l,2} .
\end{equation}
The time dependence of  the wavefunction $\psi(l,   t)  $ can be evaluated by evolving the initial state   $   \psi(l, 0) $  by the  $ U (\delta t) $ operator.  

\begin{figure}
\centering\includegraphics[width=0.95\textwidth]{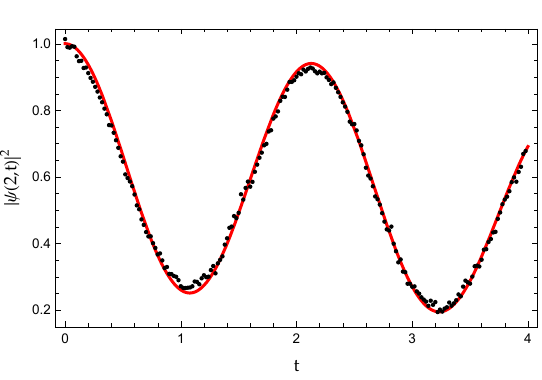}  
\caption{Demo plots of probability distribution   $ | \psi (2, t) |^2 $ on IBM hardware (black dots) without error mitigation vs. exact numerical result (red solid)  for a  2-qubit and $4$-site finite system 
under the parameters  $\Delta_a = 5 $ ,  $\Delta_b=1$, $F = 1.5 $. The time evolution step is $\delta t = 0.02$.  }
\label{psisqplot}
\end{figure}

A comparison of the exact solution of $\psi(2,   t)  $ and the results from running on IBM hardware is shown in Fig.~\ref{psisqplot} with two qubits. We see that hardware results are comparable to the exact solution even without any error mitigation, deviating only slightly at the local extrema. 

\begin{figure}
\includegraphics[width=0.95\textwidth]{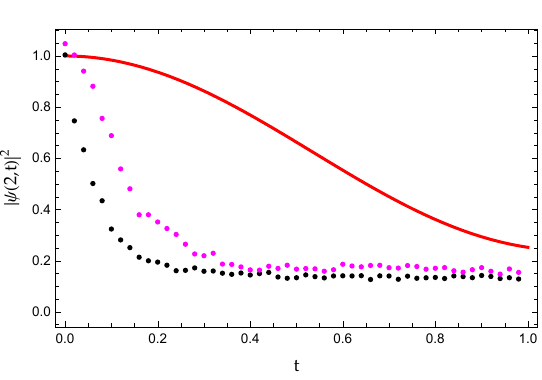}  
\caption{Similar to Fig.~\ref{psisqplot}, but for a 3-qubit and $8$-site finite system.   }
\label{psisq3qbuitplot}
\end{figure}

This is not the case for the 3-qubit results shown in Fig.~\ref{psisq3qbuitplot}. The expectation values drop off dramatically compared to the exact solution due to noise on the IBM machines. In an effort to improve the results, we attempted several error mitigation methods (provided by IBM qiskit), such as twirled Readout Error Extinction \cite{PhysRevA.105.032620},  Zero Noise Extrapolation (ZNE) \cite{PhysRevLett.119.180509}, Pauli Twirling \cite{PhysRevA.94.052325}, etc. 
Only the ZNE  method shows a significant improvement, but even so, it is still way off the  exact result.

\begin{figure*}
\frame{\includegraphics[width=.99\textwidth]{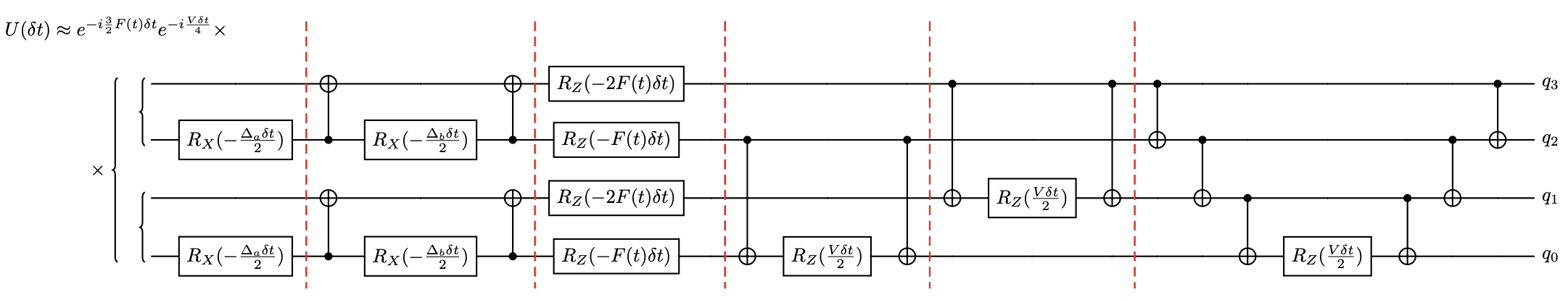}}
\caption{Quantum circuit of the lowest order $U( \delta t )$ in Eq.\eqref{U3}  for  4-qubit (two sets of 2-qubit) two interacting particles on 4-site periodic chain.} \label{Fig4qubitU}
\end{figure*}

\begin{figure}
\includegraphics[width=0.95\textwidth]{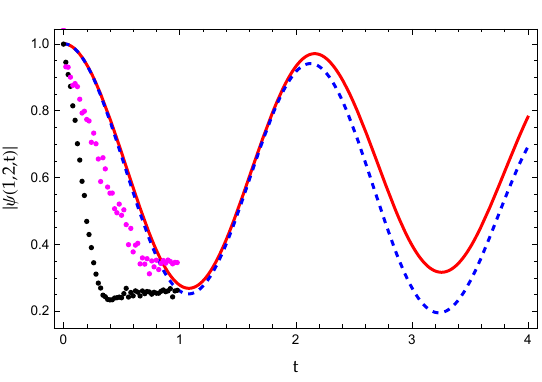}  
\caption{Demo plots of   $ | \psi (1, 2, t) | $ on IBM hardware (black dots) without error mitigation, results  with error mitigation (purple dots)  of Zero Noise Extrapolation method \cite{PhysRevLett.119.180509}   vs. exact numerical result (red solid)  for a  4-qubit (two sets 2-qubit) and $4$-site two interacting particles   system  under the parameters  $V=10$, $\Delta_a = 5 $ ,  $\Delta_b=1$, $F = 1.5 $. The time evolution step is $\delta t = 0.02$. The $ | \psi (1, 2, t) | $ with interaction turned off ($V=0$, blue dashed) is also plotted as a comparison. }
\label{interactplot}
\end{figure}

With interaction turned on, 4-qubit quantum circuit  of   lowest order Trotterization approximation of Eq.\eqref{U3} for two interacting particles on a 4-site periodic chain is shown in Fig.~\ref{Fig4qubitU}.   The initial state is chosen as a spike at site $l_1=1$ and $l_2=2$ for particle-1 and particle-2 respectively, 
\begin{equation}
\psi(l_1, l_2, 0) = \delta_{l_1,1} \delta_{l_2,2} .
\end{equation}
A comparison of the exact solution of $\psi(1,2,   t)  $ and the results from running on IBM hardware is shown in Fig.~\ref{interactplot} with four qubits. Similar to the 3-qubit non-interacting single particle result shown in Fig.~\ref{psisq3qbuitplot}, the 4-qubit two-particle interacting system is also way off the exact result even with error mitigation effort. To highlight the effects of the interaction, the  exact solution of $\psi(1,2,   t)  $ with interaction turned off ($V=0$) is also plotted in Fig.~\ref{interactplot}. At the limit of $V=0$, the two-particle wave function is simply the product of two single particle wave function: 
\begin{equation}
\psi(l_1, l_2, t) \stackrel{V \rightarrow 0}{\rightarrow } \psi(l_1,   t) \psi( l_2, t)  .
\end{equation}

\vspace{2cm}
\section{Summary and outlook}\label{sec:summary}

Instead of mapping the second quantization of the system Hamiltonian to  qubit Pauli gates representation via the Jordan-Wigner transform,   we  explored the use of statevector basis representation to simulate few-body dynamics on quantum computers in the present work.  For few-body dynamics, we demonstrated that the statevector basis representation is much more efficient on the number of quantum registers.  For a single-particle excitation state, $\Gamma$ qubits can simulate $N=2^\Gamma$ number of sites of a 1D chain system, in comparison to $N$ qubits for $N$-site 1D system via the Jordan-Wigner approach. 

The two-band diatomic tight-binding 1D model, which is relatively simple but still rich in physics content, is used to demonstrate the effectiveness of quantum simulation under statevector basis representation.   Exact expressions for quantum circuits of unitary time evolution of hopping terms, electric field term, and interaction term are obtained. The few-body dynamics of the  model can be simulated  with only a few qubits and simple quantum circuits.   
Numerical tests on IBM hardware show excellent agreement on a 2-qubit and 4-site system, even without error mitigation.
On the other hand, testing results on the same IBM hardware are not satisfactory when the number of qubits is greater than two.
Unfortunately, without significant effort on error mitigation and error correction, quantum simulation on the current noisy intermediate-scale quantum (NISQ) hardware are too noisy with more qubits.   For the interacting case, even for the two particles on 4-site with two sets of 2-qubit, the testing result on IBM hardware is not satisfactory. The reason of failure is not completely clear at this point. We suspect that the deep quantum circuit may be the primary cause. Using non-interacting 3-qubit case as an example, in terms of $U1$, $U3$ and $CX$ gates, the depth of decomposed quantum circuit $U(\delta t)$ in Fig.~\ref{Fig3qubitU} is 25 layers, with 4 $U1$ gates, 13   $U3$ gates and 14 expensive $CX$ gates. Consequently, $U(10 \delta t)$ would involve  250 layers which may become too challenging for current quantum hardware with decoherence time in the range of hundred microseconds without significant efforts on error mitigation and error correction.  

The extension to higher dimensions in the two-band tight-binding model is fairly straightforward, by adding extra sets of quantum registers for extra dimensions. This is outlined in Appendix \ref{sec:2Dextension}.  The extension to include multiple orbits per site could possibly be dealt with in a similar way to that for higher dimensions.   

The ultimate goal is to explore the quantum simulation of Bloch oscillation of $\mathcal{P T}$-symmetric systems by using complex-valued  hopping parameters in the two-band diatomic tight-binding model. Such systems possess non-Hermitian dynamics that open up new avenues in quantum field theory whose simulation on unitary gate operation quantum computers  may be accomplished via the Naimark dilation procedure,   see e.g. Refs.~\cite{doi:10.1126/science.aaw8205,Dogra2021}.

\acknowledgments
This research is supported by the U.S. National Science Foundation under grant PHY-2418937 (P.G.) and in part PHY-1748958 (P.G.), and the U.S. Department of Energy under grant  DE-FG02-95ER40907 (F.L.).

\pagebreak
\bibliography{ALL-REF.bib}

\begin{thebibliography}{32}%
\makeatletter
\providecommand \@ifxundefined [1]{%
 \@ifx{#1\undefined}
}%
\providecommand \@ifnum [1]{%
 \ifnum #1\expandafter \@firstoftwo
 \else \expandafter \@secondoftwo
 \fi
}%
\providecommand \@ifx [1]{%
 \ifx #1\expandafter \@firstoftwo
 \else \expandafter \@secondoftwo
 \fi
}%
\providecommand \natexlab [1]{#1}%
\providecommand \enquote  [1]{``#1''}%
\providecommand \bibnamefont  [1]{#1}%
\providecommand \bibfnamefont [1]{#1}%
\providecommand \citenamefont [1]{#1}%
\providecommand \href@noop [0]{\@secondoftwo}%
\providecommand \href [0]{\begingroup \@sanitize@url \@href}%
\providecommand \@href[1]{\@@startlink{#1}\@@href}%
\providecommand \@@href[1]{\endgroup#1\@@endlink}%
\providecommand \@sanitize@url [0]{\catcode `\\12\catcode `\$12\catcode
  `\&12\catcode `\#12\catcode `\^12\catcode `\_12\catcode `\%12\relax}%
\providecommand \@@startlink[1]{}%
\providecommand \@@endlink[0]{}%
\providecommand \url  [0]{\begingroup\@sanitize@url \@url }%
\providecommand \@url [1]{\endgroup\@href {#1}{\urlprefix }}%
\providecommand \urlprefix  [0]{URL }%
\providecommand \Eprint [0]{\href }%
\providecommand \doibase [0]{https://doi.org/}%
\providecommand \selectlanguage [0]{\@gobble}%
\providecommand \bibinfo  [0]{\@secondoftwo}%
\providecommand \bibfield  [0]{\@secondoftwo}%
\providecommand \translation [1]{[#1]}%
\providecommand \BibitemOpen [0]{}%
\providecommand \bibitemStop [0]{}%
\providecommand \bibitemNoStop [0]{.\EOS\space}%
\providecommand \EOS [0]{\spacefactor3000\relax}%
\providecommand \BibitemShut  [1]{\csname bibitem#1\endcsname}%
\let\auto@bib@innerbib\@empty
\bibitem [{\citenamefont {Schaefer}\ \emph {et~al.}(2011)\citenamefont
  {Schaefer}, \citenamefont {Sommer},\ and\ \citenamefont
  {Virotta}}]{SCHAEFER201193}%
  \BibitemOpen
  \bibfield  {author} {\bibinfo {author} {\bibfnamefont {S.}~\bibnamefont
  {Schaefer}}, \bibinfo {author} {\bibfnamefont {R.}~\bibnamefont {Sommer}},\
  and\ \bibinfo {author} {\bibfnamefont {F.}~\bibnamefont {Virotta}} (\bibinfo
  {collaboration} {ALPHA}),\ }\bibfield  {title} {\bibinfo {title} {{Critical
  slowing down and error analysis in lattice QCD simulations}},\ }\href
  {https://doi.org/10.1016/j.nuclphysb.2010.11.020} {\bibfield  {journal}
  {\bibinfo  {journal} {Nucl. Phys. B}\ }\textbf {\bibinfo {volume} {845}},\
  \bibinfo {pages} {93} (\bibinfo {year} {2011})},\ \Eprint
  {https://arxiv.org/abs/1009.5228} {arXiv:1009.5228 [hep-lat]} \BibitemShut
  {NoStop}%
\bibitem [{\citenamefont {Loh}\ \emph {et~al.}(1990)\citenamefont {Loh},
  \citenamefont {Gubernatis}, \citenamefont {Scalettar}, \citenamefont {White},
  \citenamefont {Scalapino},\ and\ \citenamefont {Sugar}}]{PhysRevB.41.9301}%
  \BibitemOpen
  \bibfield  {author} {\bibinfo {author} {\bibfnamefont {E.~Y.}\ \bibnamefont
  {Loh}}, \bibinfo {author} {\bibfnamefont {J.~E.}\ \bibnamefont {Gubernatis}},
  \bibinfo {author} {\bibfnamefont {R.~T.}\ \bibnamefont {Scalettar}}, \bibinfo
  {author} {\bibfnamefont {S.~R.}\ \bibnamefont {White}}, \bibinfo {author}
  {\bibfnamefont {D.~J.}\ \bibnamefont {Scalapino}},\ and\ \bibinfo {author}
  {\bibfnamefont {R.~L.}\ \bibnamefont {Sugar}},\ }\bibfield  {title} {\bibinfo
  {title} {{Sign problem in the numerical simulation of many-electron
  systems}},\ }\href {https://doi.org/10.1103/PhysRevB.41.9301} {\bibfield
  {journal} {\bibinfo  {journal} {Phys. Rev. B}\ }\textbf {\bibinfo {volume}
  {41}},\ \bibinfo {pages} {9301} (\bibinfo {year} {1990})}\BibitemShut
  {NoStop}%
\bibitem [{\citenamefont {de~Forcrand}(2009)}]{deForcrand:2009zkb}%
  \BibitemOpen
  \bibfield  {author} {\bibinfo {author} {\bibfnamefont {P.}~\bibnamefont
  {de~Forcrand}},\ }\bibfield  {title} {\bibinfo {title} {{Simulating QCD at
  finite density}},\ }\href {https://doi.org/10.22323/1.091.0010} {\bibfield
  {journal} {\bibinfo  {journal} {PoS}\ }\textbf {\bibinfo {volume}
  {LAT2009}},\ \bibinfo {pages} {010} (\bibinfo {year} {2009})},\ \Eprint
  {https://arxiv.org/abs/1005.0539} {arXiv:1005.0539 [hep-lat]} \BibitemShut
  {NoStop}%
\bibitem [{\citenamefont {Lepage}(1989)}]{lepage1989analysis}%
  \BibitemOpen
  \bibfield  {author} {\bibinfo {author} {\bibfnamefont {G.~P.}\ \bibnamefont
  {Lepage}},\ }\bibfield  {title} {\bibinfo {title} {The analysis of algorithms
  for lattice field theory},\ }\href@noop {} {\bibfield  {journal} {\bibinfo
  {journal} {Boulder ASI}\ }\textbf {\bibinfo {volume} {1989}},\ \bibinfo
  {pages} {97} (\bibinfo {year} {1989})}\BibitemShut {NoStop}%
\bibitem [{\citenamefont {Drischler}\ \emph {et~al.}(2021)\citenamefont
  {Drischler}, \citenamefont {Haxton}, \citenamefont {McElvain}, \citenamefont
  {Mereghetti}, \citenamefont {Nicholson}, \citenamefont {Vranas},\ and\
  \citenamefont {Walker-Loud}}]{DRISCHLER2021103888}%
  \BibitemOpen
  \bibfield  {author} {\bibinfo {author} {\bibfnamefont {C.}~\bibnamefont
  {Drischler}}, \bibinfo {author} {\bibfnamefont {W.}~\bibnamefont {Haxton}},
  \bibinfo {author} {\bibfnamefont {K.}~\bibnamefont {McElvain}}, \bibinfo
  {author} {\bibfnamefont {E.}~\bibnamefont {Mereghetti}}, \bibinfo {author}
  {\bibfnamefont {A.}~\bibnamefont {Nicholson}}, \bibinfo {author}
  {\bibfnamefont {P.}~\bibnamefont {Vranas}},\ and\ \bibinfo {author}
  {\bibfnamefont {A.}~\bibnamefont {Walker-Loud}},\ }\bibfield  {title}
  {\bibinfo {title} {Towards grounding nuclear physics in qcd},\ }\href
  {https://doi.org/https://doi.org/10.1016/j.ppnp.2021.103888} {\bibfield
  {journal} {\bibinfo  {journal} {Progress in Particle and Nuclear Physics}\
  }\textbf {\bibinfo {volume} {121}},\ \bibinfo {pages} {103888} (\bibinfo
  {year} {2021})}\BibitemShut {NoStop}%
\bibitem [{\citenamefont {Tacchino}\ \emph {et~al.}(2020)\citenamefont
  {Tacchino}, \citenamefont {Chiesa}, \citenamefont {Carretta},\ and\
  \citenamefont {Gerace}}]{https://doi.org/10.1002/qute.201900052}%
  \BibitemOpen
  \bibfield  {author} {\bibinfo {author} {\bibfnamefont {F.}~\bibnamefont
  {Tacchino}}, \bibinfo {author} {\bibfnamefont {A.}~\bibnamefont {Chiesa}},
  \bibinfo {author} {\bibfnamefont {S.}~\bibnamefont {Carretta}},\ and\
  \bibinfo {author} {\bibfnamefont {D.}~\bibnamefont {Gerace}},\ }\bibfield
  {title} {\bibinfo {title} {Quantum computers as universal quantum simulators:
  State-of-the-art and perspectives},\ }\href
  {https://doi.org/https://doi.org/10.1002/qute.201900052} {\bibfield
  {journal} {\bibinfo  {journal} {Advanced Quantum Technologies}\ }\textbf
  {\bibinfo {volume} {3}},\ \bibinfo {pages} {1900052} (\bibinfo {year}
  {2020})}\BibitemShut {NoStop}%
\bibitem [{\citenamefont {Hubbard}\ and\ \citenamefont
  {Flowers}(1963)}]{doi:10.1098/rspa.1963.0204}%
  \BibitemOpen
  \bibfield  {author} {\bibinfo {author} {\bibfnamefont {J.}~\bibnamefont
  {Hubbard}}\ and\ \bibinfo {author} {\bibfnamefont {B.~H.}\ \bibnamefont
  {Flowers}},\ }\bibfield  {title} {\bibinfo {title} {Electron correlations in
  narrow energy bands},\ }\href {https://doi.org/10.1098/rspa.1963.0204}
  {\bibfield  {journal} {\bibinfo  {journal} {Proceedings of the Royal Society
  of London. Series A. Mathematical and Physical Sciences}\ }\textbf {\bibinfo
  {volume} {276}},\ \bibinfo {pages} {238} (\bibinfo {year}
  {1963})}\BibitemShut {NoStop}%
\bibitem [{\citenamefont {Ortiz}\ \emph {et~al.}(2001)\citenamefont {Ortiz},
  \citenamefont {Gubernatis}, \citenamefont {Knill},\ and\ \citenamefont
  {Laflamme}}]{PhysRevA.64.022319}%
  \BibitemOpen
  \bibfield  {author} {\bibinfo {author} {\bibfnamefont {G.}~\bibnamefont
  {Ortiz}}, \bibinfo {author} {\bibfnamefont {J.~E.}\ \bibnamefont
  {Gubernatis}}, \bibinfo {author} {\bibfnamefont {E.}~\bibnamefont {Knill}},\
  and\ \bibinfo {author} {\bibfnamefont {R.}~\bibnamefont {Laflamme}},\
  }\bibfield  {title} {\bibinfo {title} {Quantum algorithms for fermionic
  simulations},\ }\href {https://doi.org/10.1103/PhysRevA.64.022319} {\bibfield
   {journal} {\bibinfo  {journal} {Phys. Rev. A}\ }\textbf {\bibinfo {volume}
  {64}},\ \bibinfo {pages} {022319} (\bibinfo {year} {2001})}\BibitemShut
  {NoStop}%
\bibitem [{\citenamefont {Somma}\ \emph {et~al.}(2002)\citenamefont {Somma},
  \citenamefont {Ortiz}, \citenamefont {Gubernatis}, \citenamefont {Knill},\
  and\ \citenamefont {Laflamme}}]{PhysRevA.65.042323}%
  \BibitemOpen
  \bibfield  {author} {\bibinfo {author} {\bibfnamefont {R.}~\bibnamefont
  {Somma}}, \bibinfo {author} {\bibfnamefont {G.}~\bibnamefont {Ortiz}},
  \bibinfo {author} {\bibfnamefont {J.~E.}\ \bibnamefont {Gubernatis}},
  \bibinfo {author} {\bibfnamefont {E.}~\bibnamefont {Knill}},\ and\ \bibinfo
  {author} {\bibfnamefont {R.}~\bibnamefont {Laflamme}},\ }\bibfield  {title}
  {\bibinfo {title} {Simulating physical phenomena by quantum networks},\
  }\href {https://doi.org/10.1103/PhysRevA.65.042323} {\bibfield  {journal}
  {\bibinfo  {journal} {Phys. Rev. A}\ }\textbf {\bibinfo {volume} {65}},\
  \bibinfo {pages} {042323} (\bibinfo {year} {2002})}\BibitemShut {NoStop}%
\bibitem [{\citenamefont {Jordan}\ and\ \citenamefont
  {Wigner}(1928)}]{JordanWigner}%
  \BibitemOpen
  \bibfield  {author} {\bibinfo {author} {\bibfnamefont {P.}~\bibnamefont
  {Jordan}}\ and\ \bibinfo {author} {\bibfnamefont {E.}~\bibnamefont
  {Wigner}},\ }\bibfield  {title} {\bibinfo {title} {{\"U}ber das paulische
  {\"a}quivalenzverbot},\ }\href {https://doi.org/10.1007/BF01331938}
  {\bibfield  {journal} {\bibinfo  {journal} {Zeitschrift f{\"u}r Physik}\
  }\textbf {\bibinfo {volume} {47}},\ \bibinfo {pages} {631} (\bibinfo {year}
  {1928})}\BibitemShut {NoStop}%
\bibitem [{\citenamefont {Trotter}(1959)}]{TrotterH.F.1959Otpo}%
  \BibitemOpen
  \bibfield  {author} {\bibinfo {author} {\bibfnamefont {H.~F.}\ \bibnamefont
  {Trotter}},\ }\bibfield  {title} {\bibinfo {title} {On the product of
  semi-groups of operators},\ }\href {http://www.jstor.org/stable/2033649}
  {\bibfield  {journal} {\bibinfo  {journal} {Proceedings of the American
  Mathematical Society}\ }\textbf {\bibinfo {volume} {10}},\ \bibinfo {pages}
  {545} (\bibinfo {year} {1959})}\BibitemShut {NoStop}%
\bibitem [{\citenamefont {Hatano}\ and\ \citenamefont
  {Suzuki}(2005)}]{Hatano:2005gh}%
  \BibitemOpen
  \bibfield  {author} {\bibinfo {author} {\bibfnamefont {N.}~\bibnamefont
  {Hatano}}\ and\ \bibinfo {author} {\bibfnamefont {M.}~\bibnamefont
  {Suzuki}},\ }\bibfield  {title} {\bibinfo {title} {{Finding Exponential
  Product Formulas of Higher Orders}},\ }\href
  {https://doi.org/10.1007/11526216_2} {\bibfield  {journal} {\bibinfo
  {journal} {Lect. Notes Phys.}\ }\textbf {\bibinfo {volume} {679}},\ \bibinfo
  {pages} {37} (\bibinfo {year} {2005})},\ \Eprint
  {https://arxiv.org/abs/math-ph/0506007} {arXiv:math-ph/0506007} \BibitemShut
  {NoStop}%
\bibitem [{\citenamefont {Bloch}(1929)}]{Bloch1929}%
  \BibitemOpen
  \bibfield  {author} {\bibinfo {author} {\bibfnamefont {F.}~\bibnamefont
  {Bloch}},\ }\bibfield  {title} {\bibinfo {title} {{\"U}ber die
  quantenmechanik der elektronen in kristallgittern},\ }\href
  {https://doi.org/10.1007/BF01339455} {\bibfield  {journal} {\bibinfo
  {journal} {Zeitschrift f{\"u}r Physik}\ }\textbf {\bibinfo {volume} {52}},\
  \bibinfo {pages} {555} (\bibinfo {year} {1929})}\BibitemShut {NoStop}%
\bibitem [{\citenamefont {Zener}\ and\ \citenamefont
  {Fowler}(1934)}]{doi:10.1098/rspa.1934.0116}%
  \BibitemOpen
  \bibfield  {author} {\bibinfo {author} {\bibfnamefont {C.}~\bibnamefont
  {Zener}}\ and\ \bibinfo {author} {\bibfnamefont {R.~H.}\ \bibnamefont
  {Fowler}},\ }\bibfield  {title} {\bibinfo {title} {A theory of the electrical
  breakdown of solid dielectrics},\ }\href
  {https://doi.org/10.1098/rspa.1934.0116} {\bibfield  {journal} {\bibinfo
  {journal} {Proceedings of the Royal Society of London. Series A, Containing
  Papers of a Mathematical and Physical Character}\ }\textbf {\bibinfo {volume}
  {145}},\ \bibinfo {pages} {523} (\bibinfo {year} {1934})}\BibitemShut
  {NoStop}%
\bibitem [{\citenamefont {Wannier}(1962)}]{RevModPhys.34.645}%
  \BibitemOpen
  \bibfield  {author} {\bibinfo {author} {\bibfnamefont {G.~H.}\ \bibnamefont
  {Wannier}},\ }\bibfield  {title} {\bibinfo {title} {Dynamics of band
  electrons in electric and magnetic fields},\ }\href
  {https://doi.org/10.1103/RevModPhys.34.645} {\bibfield  {journal} {\bibinfo
  {journal} {Rev. Mod. Phys.}\ }\textbf {\bibinfo {volume} {34}},\ \bibinfo
  {pages} {645} (\bibinfo {year} {1962})}\BibitemShut {NoStop}%
\bibitem [{\citenamefont {Hartmann}\ \emph {et~al.}(2004)\citenamefont
  {Hartmann}, \citenamefont {Keck}, \citenamefont {Korsch},\ and\ \citenamefont
  {Mossmann}}]{Hartmann_2004}%
  \BibitemOpen
  \bibfield  {author} {\bibinfo {author} {\bibfnamefont {T.}~\bibnamefont
  {Hartmann}}, \bibinfo {author} {\bibfnamefont {F.}~\bibnamefont {Keck}},
  \bibinfo {author} {\bibfnamefont {H.~J.}\ \bibnamefont {Korsch}},\ and\
  \bibinfo {author} {\bibfnamefont {S.}~\bibnamefont {Mossmann}},\ }\bibfield
  {title} {\bibinfo {title} {Dynamics of bloch oscillations},\ }\href
  {https://doi.org/10.1088/1367-2630/6/1/002} {\bibfield  {journal} {\bibinfo
  {journal} {New Journal of Physics}\ }\textbf {\bibinfo {volume} {6}},\
  \bibinfo {pages} {2} (\bibinfo {year} {2004})}\BibitemShut {NoStop}%
\bibitem [{\citenamefont {Callaway}(1974)}]{callaway1974quantum}%
  \BibitemOpen
  \bibfield  {author} {\bibinfo {author} {\bibfnamefont {J.}~\bibnamefont
  {Callaway}},\ }\href {https://doi.org/10.1016/C2009-0-22300-2} {\emph
  {\bibinfo {title} {Quantum Theory of the Solid State}}},\ Quantum Theory of
  the Solid State\ (\bibinfo  {publisher} {Academic Press},\ \bibinfo {year}
  {1974})\BibitemShut {NoStop}%
\bibitem [{\citenamefont {Glück}\ \emph {et~al.}(2002)\citenamefont {Glück},
  \citenamefont {{R. Kolovsky}},\ and\ \citenamefont {Korsch}}]{GLUCK2002103}%
  \BibitemOpen
  \bibfield  {author} {\bibinfo {author} {\bibfnamefont {M.}~\bibnamefont
  {Glück}}, \bibinfo {author} {\bibfnamefont {A.}~\bibnamefont {{R.
  Kolovsky}}},\ and\ \bibinfo {author} {\bibfnamefont {H.~J.}\ \bibnamefont
  {Korsch}},\ }\bibfield  {title} {\bibinfo {title} {Wannier–stark resonances
  in optical and semiconductor superlattices},\ }\href
  {https://doi.org/https://doi.org/10.1016/S0370-1573(02)00142-4} {\bibfield
  {journal} {\bibinfo  {journal} {Physics Reports}\ }\textbf {\bibinfo {volume}
  {366}},\ \bibinfo {pages} {103} (\bibinfo {year} {2002})}\BibitemShut
  {NoStop}%
\bibitem [{\citenamefont {Feldmann}\ \emph {et~al.}(1992)\citenamefont
  {Feldmann}, \citenamefont {Leo}, \citenamefont {Shah}, \citenamefont
  {Miller}, \citenamefont {Cunningham}, \citenamefont {Meier}, \citenamefont
  {von Plessen}, \citenamefont {Schulze}, \citenamefont {Thomas},\ and\
  \citenamefont {Schmitt-Rink}}]{PhysRevB.46.7252}%
  \BibitemOpen
  \bibfield  {author} {\bibinfo {author} {\bibfnamefont {J.}~\bibnamefont
  {Feldmann}}, \bibinfo {author} {\bibfnamefont {K.}~\bibnamefont {Leo}},
  \bibinfo {author} {\bibfnamefont {J.}~\bibnamefont {Shah}}, \bibinfo {author}
  {\bibfnamefont {D.~A.~B.}\ \bibnamefont {Miller}}, \bibinfo {author}
  {\bibfnamefont {J.~E.}\ \bibnamefont {Cunningham}}, \bibinfo {author}
  {\bibfnamefont {T.}~\bibnamefont {Meier}}, \bibinfo {author} {\bibfnamefont
  {G.}~\bibnamefont {von Plessen}}, \bibinfo {author} {\bibfnamefont
  {A.}~\bibnamefont {Schulze}}, \bibinfo {author} {\bibfnamefont
  {P.}~\bibnamefont {Thomas}},\ and\ \bibinfo {author} {\bibfnamefont
  {S.}~\bibnamefont {Schmitt-Rink}},\ }\bibfield  {title} {\bibinfo {title}
  {Optical investigation of bloch oscillations in a semiconductor
  superlattice},\ }\href {https://doi.org/10.1103/PhysRevB.46.7252} {\bibfield
  {journal} {\bibinfo  {journal} {Phys. Rev. B}\ }\textbf {\bibinfo {volume}
  {46}},\ \bibinfo {pages} {7252} (\bibinfo {year} {1992})}\BibitemShut
  {NoStop}%
\bibitem [{\citenamefont {Ben~Dahan}\ \emph {et~al.}(1996)\citenamefont
  {Ben~Dahan}, \citenamefont {Peik}, \citenamefont {Reichel}, \citenamefont
  {Castin},\ and\ \citenamefont {Salomon}}]{PhysRevLett.76.4508}%
  \BibitemOpen
  \bibfield  {author} {\bibinfo {author} {\bibfnamefont {M.}~\bibnamefont
  {Ben~Dahan}}, \bibinfo {author} {\bibfnamefont {E.}~\bibnamefont {Peik}},
  \bibinfo {author} {\bibfnamefont {J.}~\bibnamefont {Reichel}}, \bibinfo
  {author} {\bibfnamefont {Y.}~\bibnamefont {Castin}},\ and\ \bibinfo {author}
  {\bibfnamefont {C.}~\bibnamefont {Salomon}},\ }\bibfield  {title} {\bibinfo
  {title} {Bloch oscillations of atoms in an optical potential},\ }\href
  {https://doi.org/10.1103/PhysRevLett.76.4508} {\bibfield  {journal} {\bibinfo
   {journal} {Phys. Rev. Lett.}\ }\textbf {\bibinfo {volume} {76}},\ \bibinfo
  {pages} {4508} (\bibinfo {year} {1996})}\BibitemShut {NoStop}%
\bibitem [{\citenamefont {Pertsch}\ \emph {et~al.}(1999)\citenamefont
  {Pertsch}, \citenamefont {Dannberg}, \citenamefont {Elflein}, \citenamefont
  {Br\"auer},\ and\ \citenamefont {Lederer}}]{PhysRevLett.83.4752}%
  \BibitemOpen
  \bibfield  {author} {\bibinfo {author} {\bibfnamefont {T.}~\bibnamefont
  {Pertsch}}, \bibinfo {author} {\bibfnamefont {P.}~\bibnamefont {Dannberg}},
  \bibinfo {author} {\bibfnamefont {W.}~\bibnamefont {Elflein}}, \bibinfo
  {author} {\bibfnamefont {A.}~\bibnamefont {Br\"auer}},\ and\ \bibinfo
  {author} {\bibfnamefont {F.}~\bibnamefont {Lederer}},\ }\bibfield  {title}
  {\bibinfo {title} {Optical bloch oscillations in temperature tuned waveguide
  arrays},\ }\href {https://doi.org/10.1103/PhysRevLett.83.4752} {\bibfield
  {journal} {\bibinfo  {journal} {Phys. Rev. Lett.}\ }\textbf {\bibinfo
  {volume} {83}},\ \bibinfo {pages} {4752} (\bibinfo {year}
  {1999})}\BibitemShut {NoStop}%
\bibitem [{\citenamefont {Morandotti}\ \emph {et~al.}(1999)\citenamefont
  {Morandotti}, \citenamefont {Peschel}, \citenamefont {Aitchison},
  \citenamefont {Eisenberg},\ and\ \citenamefont
  {Silberberg}}]{PhysRevLett.83.4756}%
  \BibitemOpen
  \bibfield  {author} {\bibinfo {author} {\bibfnamefont {R.}~\bibnamefont
  {Morandotti}}, \bibinfo {author} {\bibfnamefont {U.}~\bibnamefont {Peschel}},
  \bibinfo {author} {\bibfnamefont {J.~S.}\ \bibnamefont {Aitchison}}, \bibinfo
  {author} {\bibfnamefont {H.~S.}\ \bibnamefont {Eisenberg}},\ and\ \bibinfo
  {author} {\bibfnamefont {Y.}~\bibnamefont {Silberberg}},\ }\bibfield  {title}
  {\bibinfo {title} {Experimental observation of linear and nonlinear optical
  bloch oscillations},\ }\href {https://doi.org/10.1103/PhysRevLett.83.4756}
  {\bibfield  {journal} {\bibinfo  {journal} {Phys. Rev. Lett.}\ }\textbf
  {\bibinfo {volume} {83}},\ \bibinfo {pages} {4756} (\bibinfo {year}
  {1999})}\BibitemShut {NoStop}%
\bibitem [{\citenamefont {Song}\ \emph {et~al.}(2024)\citenamefont {Song},
  \citenamefont {Xiang}, \citenamefont {Zhang}, \citenamefont {Wang},
  \citenamefont {Guo}, \citenamefont {Ruan}, \citenamefont {Song},
  \citenamefont {Xu}, \citenamefont {Gao}, \citenamefont {Fan},\ and\
  \citenamefont {Zheng}}]{SongPRX2024}%
  \BibitemOpen
  \bibfield  {author} {\bibinfo {author} {\bibfnamefont {P.}~\bibnamefont
  {Song}}, \bibinfo {author} {\bibfnamefont {Z.}~\bibnamefont {Xiang}},
  \bibinfo {author} {\bibfnamefont {Y.-X.}\ \bibnamefont {Zhang}}, \bibinfo
  {author} {\bibfnamefont {Z.}~\bibnamefont {Wang}}, \bibinfo {author}
  {\bibfnamefont {X.}~\bibnamefont {Guo}}, \bibinfo {author} {\bibfnamefont
  {X.}~\bibnamefont {Ruan}}, \bibinfo {author} {\bibfnamefont {X.}~\bibnamefont
  {Song}}, \bibinfo {author} {\bibfnamefont {K.}~\bibnamefont {Xu}}, \bibinfo
  {author} {\bibfnamefont {Y.~Y.}\ \bibnamefont {Gao}}, \bibinfo {author}
  {\bibfnamefont {H.}~\bibnamefont {Fan}},\ and\ \bibinfo {author}
  {\bibfnamefont {D.}~\bibnamefont {Zheng}},\ }\bibfield  {title} {\bibinfo
  {title} {Coherent control of bloch oscillations in a superconducting
  circuit},\ }\href {https://doi.org/10.1103/PRXQuantum.5.020302} {\bibfield
  {journal} {\bibinfo  {journal} {PRX Quantum}\ }\textbf {\bibinfo {volume}
  {5}},\ \bibinfo {pages} {020302} (\bibinfo {year} {2024})}\BibitemShut
  {NoStop}%
\bibitem [{\citenamefont {Guo}\ \emph {et~al.}(2021)\citenamefont {Guo},
  \citenamefont {Ge}, \citenamefont {Li}, \citenamefont {Wang}, \citenamefont
  {Zhang}, \citenamefont {Song}, \citenamefont {Xiang}, \citenamefont {Song},
  \citenamefont {Jin}, \citenamefont {Lu}, \citenamefont {Xu}, \citenamefont
  {Zheng},\ and\ \citenamefont {Fan}}]{GuoNature2021}%
  \BibitemOpen
  \bibfield  {author} {\bibinfo {author} {\bibfnamefont {X.-Y.}\ \bibnamefont
  {Guo}}, \bibinfo {author} {\bibfnamefont {Z.-Y.}\ \bibnamefont {Ge}},
  \bibinfo {author} {\bibfnamefont {H.}~\bibnamefont {Li}}, \bibinfo {author}
  {\bibfnamefont {Z.}~\bibnamefont {Wang}}, \bibinfo {author} {\bibfnamefont
  {Y.-R.}\ \bibnamefont {Zhang}}, \bibinfo {author} {\bibfnamefont
  {P.}~\bibnamefont {Song}}, \bibinfo {author} {\bibfnamefont {Z.}~\bibnamefont
  {Xiang}}, \bibinfo {author} {\bibfnamefont {X.}~\bibnamefont {Song}},
  \bibinfo {author} {\bibfnamefont {Y.}~\bibnamefont {Jin}}, \bibinfo {author}
  {\bibfnamefont {L.}~\bibnamefont {Lu}}, \bibinfo {author} {\bibfnamefont
  {K.}~\bibnamefont {Xu}}, \bibinfo {author} {\bibfnamefont {D.}~\bibnamefont
  {Zheng}},\ and\ \bibinfo {author} {\bibfnamefont {H.}~\bibnamefont {Fan}},\
  }\bibfield  {title} {\bibinfo {title} {Observation of bloch oscillations and
  wannier-stark localization on a superconducting quantum processor},\ }\href
  {https://doi.org/10.1038/s41534-021-00385-3} {\bibfield  {journal} {\bibinfo
  {journal} {npj Quantum Information}\ }\textbf {\bibinfo {volume} {7}},\
  \bibinfo {pages} {51} (\bibinfo {year} {2021})}\BibitemShut {NoStop}%
\bibitem [{\citenamefont {Holthaus}\ and\ \citenamefont
  {and}(1996)}]{Holthaus01081996}%
  \BibitemOpen
  \bibfield  {author} {\bibinfo {author} {\bibfnamefont {M.}~\bibnamefont
  {Holthaus}}\ and\ \bibinfo {author} {\bibfnamefont {D.~W.~H.}\ \bibnamefont
  {and}},\ }\bibfield  {title} {\bibinfo {title} {Localization effects in
  ac-driven tight-binding lattices},\ }\href
  {https://doi.org/10.1080/01418639608240331} {\bibfield  {journal} {\bibinfo
  {journal} {Philosophical Magazine B}\ }\textbf {\bibinfo {volume} {74}},\
  \bibinfo {pages} {105} (\bibinfo {year} {1996})},\ \Eprint
  {https://arxiv.org/abs/https://doi.org/10.1080/01418639608240331}
  {https://doi.org/10.1080/01418639608240331} \BibitemShut {NoStop}%
\bibitem [{\citenamefont {Guo}(2025)}]{Guo:2025vgk}%
  \BibitemOpen
  \bibfield  {author} {\bibinfo {author} {\bibfnamefont {P.}~\bibnamefont
  {Guo}},\ }\bibfield  {title} {\bibinfo {title} {{Toward extracting scattering
  phase shift from integrated correlation functions on quantum computers}},\
  }\href@noop {} {\bibfield  {journal} {\bibinfo  {journal} {arXiv preprint}\ }
  (\bibinfo {year} {2025})},\ \Eprint {https://arxiv.org/abs/2504.14474}
  {arXiv:2504.14474 [quant-ph]} \BibitemShut {NoStop}%
\bibitem [{\citenamefont {van~den Berg}\ \emph {et~al.}(2022)\citenamefont
  {van~den Berg}, \citenamefont {Minev},\ and\ \citenamefont
  {Temme}}]{PhysRevA.105.032620}%
  \BibitemOpen
  \bibfield  {author} {\bibinfo {author} {\bibfnamefont {E.}~\bibnamefont
  {van~den Berg}}, \bibinfo {author} {\bibfnamefont {Z.~K.}\ \bibnamefont
  {Minev}},\ and\ \bibinfo {author} {\bibfnamefont {K.}~\bibnamefont {Temme}},\
  }\bibfield  {title} {\bibinfo {title} {Model-free readout-error mitigation
  for quantum expectation values},\ }\href
  {https://doi.org/10.1103/PhysRevA.105.032620} {\bibfield  {journal} {\bibinfo
   {journal} {Phys. Rev. A}\ }\textbf {\bibinfo {volume} {105}},\ \bibinfo
  {pages} {032620} (\bibinfo {year} {2022})}\BibitemShut {NoStop}%
\bibitem [{\citenamefont {Temme}\ \emph {et~al.}(2017)\citenamefont {Temme},
  \citenamefont {Bravyi},\ and\ \citenamefont
  {Gambetta}}]{PhysRevLett.119.180509}%
  \BibitemOpen
  \bibfield  {author} {\bibinfo {author} {\bibfnamefont {K.}~\bibnamefont
  {Temme}}, \bibinfo {author} {\bibfnamefont {S.}~\bibnamefont {Bravyi}},\ and\
  \bibinfo {author} {\bibfnamefont {J.~M.}\ \bibnamefont {Gambetta}},\
  }\bibfield  {title} {\bibinfo {title} {Error mitigation for short-depth
  quantum circuits},\ }\href {https://doi.org/10.1103/PhysRevLett.119.180509}
  {\bibfield  {journal} {\bibinfo  {journal} {Phys. Rev. Lett.}\ }\textbf
  {\bibinfo {volume} {119}},\ \bibinfo {pages} {180509} (\bibinfo {year}
  {2017})}\BibitemShut {NoStop}%
\bibitem [{\citenamefont {Wallman}\ and\ \citenamefont
  {Emerson}(2016)}]{PhysRevA.94.052325}%
  \BibitemOpen
  \bibfield  {author} {\bibinfo {author} {\bibfnamefont {J.~J.}\ \bibnamefont
  {Wallman}}\ and\ \bibinfo {author} {\bibfnamefont {J.}~\bibnamefont
  {Emerson}},\ }\bibfield  {title} {\bibinfo {title} {Noise tailoring for
  scalable quantum computation via randomized compiling},\ }\href
  {https://doi.org/10.1103/PhysRevA.94.052325} {\bibfield  {journal} {\bibinfo
  {journal} {Phys. Rev. A}\ }\textbf {\bibinfo {volume} {94}},\ \bibinfo
  {pages} {052325} (\bibinfo {year} {2016})}\BibitemShut {NoStop}%
\bibitem [{\citenamefont {Wu}\ \emph {et~al.}(2019)\citenamefont {Wu},
  \citenamefont {Liu}, \citenamefont {Geng}, \citenamefont {Song},
  \citenamefont {Ye}, \citenamefont {Duan}, \citenamefont {Rong},\ and\
  \citenamefont {Du}}]{doi:10.1126/science.aaw8205}%
  \BibitemOpen
  \bibfield  {author} {\bibinfo {author} {\bibfnamefont {Y.}~\bibnamefont
  {Wu}}, \bibinfo {author} {\bibfnamefont {W.}~\bibnamefont {Liu}}, \bibinfo
  {author} {\bibfnamefont {J.}~\bibnamefont {Geng}}, \bibinfo {author}
  {\bibfnamefont {X.}~\bibnamefont {Song}}, \bibinfo {author} {\bibfnamefont
  {X.}~\bibnamefont {Ye}}, \bibinfo {author} {\bibfnamefont {C.-K.}\
  \bibnamefont {Duan}}, \bibinfo {author} {\bibfnamefont {X.}~\bibnamefont
  {Rong}},\ and\ \bibinfo {author} {\bibfnamefont {J.}~\bibnamefont {Du}},\
  }\bibfield  {title} {\bibinfo {title} {Observation of parity-time symmetry
  breaking in a single-spin system},\ }\href
  {https://doi.org/10.1126/science.aaw8205} {\bibfield  {journal} {\bibinfo
  {journal} {Science}\ }\textbf {\bibinfo {volume} {364}},\ \bibinfo {pages}
  {878} (\bibinfo {year} {2019})},\ \Eprint
  {https://arxiv.org/abs/https://www.science.org/doi/pdf/10.1126/science.aaw8205}
  {https://www.science.org/doi/pdf/10.1126/science.aaw8205} \BibitemShut
  {NoStop}%
\bibitem [{\citenamefont {Dogra}\ \emph {et~al.}(2021)\citenamefont {Dogra},
  \citenamefont {Melnikov},\ and\ \citenamefont {Paraoanu}}]{Dogra2021}%
  \BibitemOpen
  \bibfield  {author} {\bibinfo {author} {\bibfnamefont {S.}~\bibnamefont
  {Dogra}}, \bibinfo {author} {\bibfnamefont {A.~A.}\ \bibnamefont
  {Melnikov}},\ and\ \bibinfo {author} {\bibfnamefont {G.~S.}\ \bibnamefont
  {Paraoanu}},\ }\bibfield  {title} {\bibinfo {title} {Quantum simulation of
  parity--time symmetry breaking with a superconducting quantum processor},\
  }\href {https://doi.org/10.1038/s42005-021-00534-2} {\bibfield  {journal}
  {\bibinfo  {journal} {Communications Physics}\ }\textbf {\bibinfo {volume}
  {4}},\ \bibinfo {pages} {26} (\bibinfo {year} {2021})}\BibitemShut {NoStop}%
\bibitem [{\citenamefont {Bouchard}\ and\ \citenamefont
  {Luban}(1995)}]{PhysRevB.52.5105}%
  \BibitemOpen
  \bibfield  {author} {\bibinfo {author} {\bibfnamefont {A.~M.}\ \bibnamefont
  {Bouchard}}\ and\ \bibinfo {author} {\bibfnamefont {M.}~\bibnamefont
  {Luban}},\ }\bibfield  {title} {\bibinfo {title} {Bloch oscillations and
  other dynamical phenomena of electrons in semiconductor superlattices},\
  }\href {https://doi.org/10.1103/PhysRevB.52.5105} {\bibfield  {journal}
  {\bibinfo  {journal} {Phys. Rev. B}\ }\textbf {\bibinfo {volume} {52}},\
  \bibinfo {pages} {5105} (\bibinfo {year} {1995})}\BibitemShut {NoStop}%
\end{thebibliography}%

\pagebreak
 \appendix

\section{Basic properties of diatomic non-interacting tight-binding model}\label{sec:basicproperties}

In the absence of particle interactions and electric field, the dispersion relation of free single-particle excitation for an infinitely long chain ($N \rightarrow \infty$)  display two-band structure: 
\begin{equation}
\epsilon_{\pm} (k) = \pm \frac{1}{4} \sqrt{\Delta_a^2 + \Delta_b^2 + 2 \Delta_a \Delta_b \cos (2 k )}, \label{epsilonpm}
\end{equation}
where quasi-momentum $k$ is defined in the first Brillouin zone, $k \in [- \frac{ \pi }{2},  \frac{\pi}{2} ]$.  It is the eigenvalue solution to the free Hamiltonian in Eq.\eqref{H0twobandTB}.  The separation between two neighboring sites is set to unit one. 
 The spatial periodicity of diatomic 1D chain is thus two units.  In the limit of $\Delta_a = \Delta_b = \Delta$, the two bands merge into a single band: $\epsilon_{\pm} (k) \rightarrow - \frac{  \Delta }{2} \cos (k )$. 
 
 In general, the two-band model in Eq.(\ref{diffeqtwoband}) does not  have analytic solutions, except for the limiting case $\Delta_a = \Delta_b$ and  a constant electric field $F(t)=F$, see e.g. Ref.~\cite{Hartmann_2004,PhysRevB.52.5105}. In this case, the analytic solution is given by,
\begin{equation}
\psi(l,t) = \sum_{l' = - \infty}^{\infty} U(l, l' ; t)  \psi(l',0) ,
\end{equation}
where the unitary operator,
\begin{equation}
U(l, l' ; t) = i^{l-l'} J_{l-l'} \left (\frac{\Delta}{F} \sin \left ( \frac{F t}{2} \right ) \right) e^{- i (l+l') \frac{F t}{2}},
\end{equation}
is expressed in terms of  Bessel function of the first kind $J_n(x)$.
The expectation value of electron position at site-$l$ as a function of time is given by
\begin{equation}
 \langle  l (t) \rangle   
 = \langle  l (0) \rangle - |S_0 | \frac{\Delta}{2 F} \left [ \cos ( \theta_0 ) - \cos (F t + \theta_0 ) \right ],
\end{equation}
where 
\begin{equation}
S_0  = |S_0| e^{i \theta_0 } = \sum_{l = - \infty}^{\infty} \psi^*(l +1,0) \psi(l,0).
\end{equation}
These results clearly demonstrate  that the expectation value of site-$l$ oscillates with Bloch oscillation period $\tau_B = \frac{2\pi}{F}$ which is inversely proportional to the field strength F. (We use a natural unit system with $\hbar=c=e=1$ and unit lattice spacing).

In cases of unequal couplings $\Delta_a \neq \Delta_b $,  the wavefunction can be expanded in terms of zero-electric-field solutions  \cite{callaway1974quantum},
\begin{equation}
\psi(l, t) = \sum_{\nu = \pm }  \int_{-\frac{\pi}{2}}^{\frac{\pi}{2}} d k\, c^{(\nu)}_k ( t) \,\phi^{(\nu)}_k ( l) ,
\end{equation}
where $\phi^{(\nu)}_k ( l)$ is the eigensolution of free particle Hamiltonian with eigen-energy of $\epsilon_{\nu} (k)$ in Eq.(\ref{epsilonpm}),
\begin{equation}
\hat{H}_0  | \Phi^{(\nu)}_k \rangle  = \epsilon_{\nu} (k) | \Phi^{(\nu)}_k \rangle
\text{ with } | \Phi^{(\nu)}_k \rangle = \sum_{l=0}^{N-1}  \phi^{(\nu)}_k ( l)  | l \rangle. 
\end{equation}
The analytic expression of $ \phi^{(\nu)}_k ( l) $ is given by
\begin{align}
 \phi^{(\nu)}_k ( l)  
& =
  \begin{cases} 
   \frac{1}{\sqrt{2}} D^{(\nu)} (k) e^{- i k  l } , & l =2 n, \nonumber \\
   \frac{1}{\sqrt{2}}  e^{- i k  l  } , & l =2 n +1 ,
  \end{cases} \\
  & \text{ with } D^{(\nu)} (k)   =- \frac{4 \epsilon_{\nu} (k)}{ \Delta_a e^{i k } + \Delta_b e^{- i k}}. 
\end{align}
The expansion coefficients $c^{(\nu)}_k ( t)$ satisfy coupled differential equations,
\begin{align}
& \left [ \frac{1}{F(t)} \frac{\partial }{\partial t} +  \frac{\partial }{\partial k} \right ]  c^{(\nu)}_k ( t)  \nonumber \\
& =  i \sum_{\nu' = \pm }  \left [ X_{\nu \nu'} (k) - \delta_{\nu, \nu'} \frac{\epsilon_{\nu} (k)}{F(t)} \right ]  c^{(\nu')}_k ( t) , \label{BScoefeq}
\end{align}
where
\begin{equation}
X_{\nu \nu'} (k)  = \frac{i}{2} D^{(\nu) *} (k)  \frac{\partial }{\partial k}   D^{(\nu')} (k)  .
\end{equation}
For constant uniform electric field $F(t) = F$, and when the off-diagonal terms of $X_{\nu \nu'} (k) $ that describe the tunneling effect  are ignored, the eigenenergies are given by Wannier-Stark ladders \cite{Hartmann_2004,callaway1974quantum},
\begin{equation}
E^{(\nu)}_\alpha  \simeq  2 F \alpha + \frac{1}{\pi} \int_{-\frac{\pi}{2}}^{\frac{\pi}{2}} d p \left [ \epsilon_{\nu} (p) - F X_{\nu \nu} (p) \right ], \ \  \alpha \in \mathbb{Z} .
\end{equation}
We see that the slope of the ladder is controlled by the strength of the electric field $F$.

\begin{figure}
\includegraphics[width=0.9\textwidth]{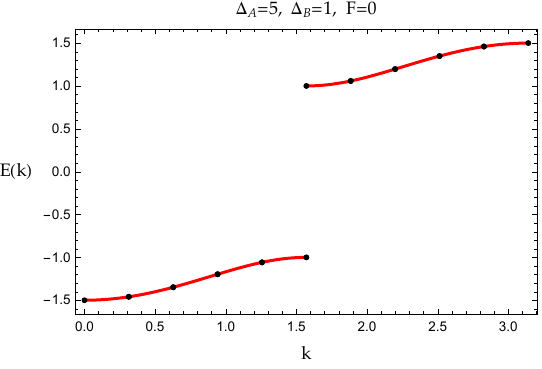}   
\includegraphics[width=0.9\textwidth]{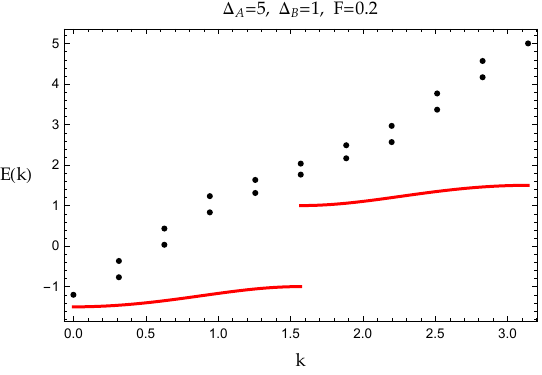}  
\includegraphics[width=0.9\textwidth]{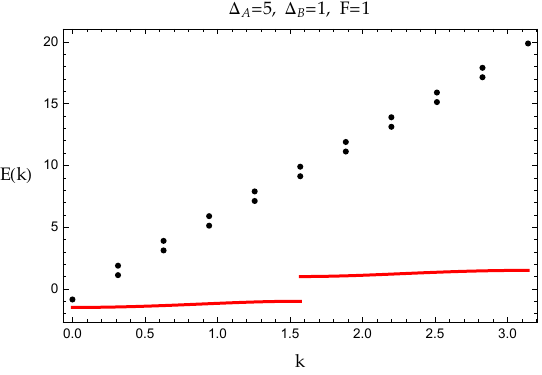}  
\caption{Dispersion relation of the electron in  the diatomic tight-binding model $\hat{H}_{sv} (t)$  in Eq.(\ref{Hftmatrixrep})  as a function of $k = \frac{\pi}{N}  [0, \frac{N}{2}-1]$ for a $N=20$-site system. The model has different hopping parameters $\Delta_a=5$, $\Delta_b=1$, no electron-electron interactions, and a constant electric field at three magnitudes $F = 0, 0.2, 1$ (top to bottom). For comparison, the zero-electric-field dispersion relation (red solid curves) in Eq.\eqref{epsilonpm} are also plotted. }
\label{Ekplots}
\end{figure}

\begin{figure}
\includegraphics[width=0.9\textwidth]{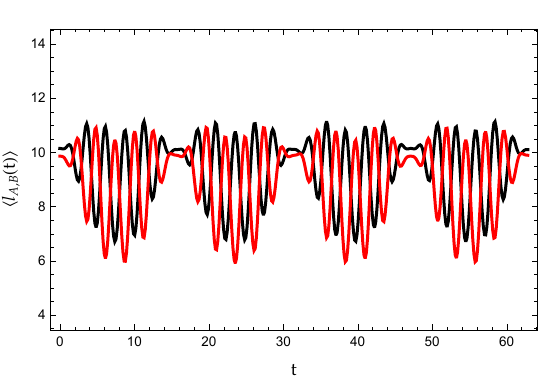}   
\includegraphics[width=0.9\textwidth]{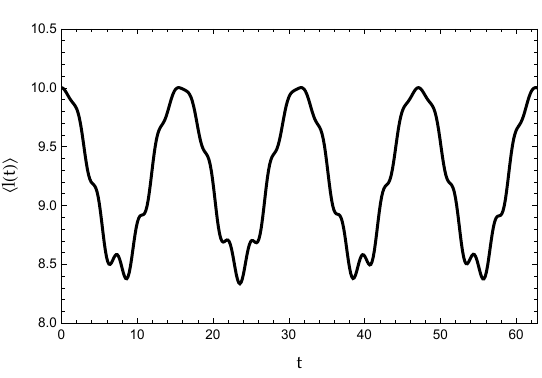}  
\caption{Expectation value of electron position in a $20$-site finite system with hopping parameters $\Delta_a=5$, $\Delta_b=1$ and constant electric field $F =  0.2$. Top:  type-A  (black)  vs. type-B  (red) as defined in Eq.\eqref{lab}. Bottom: both types as an average defined in Eq.\eqref{lave}. }
\label{ztplot0s}
\end{figure}

\begin{figure}
\includegraphics[width=0.9\textwidth]{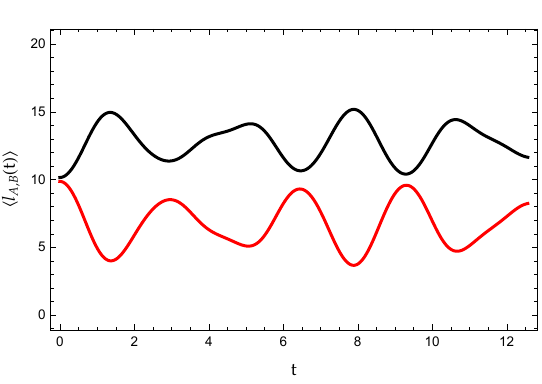}   
\includegraphics[width=0.9\textwidth]{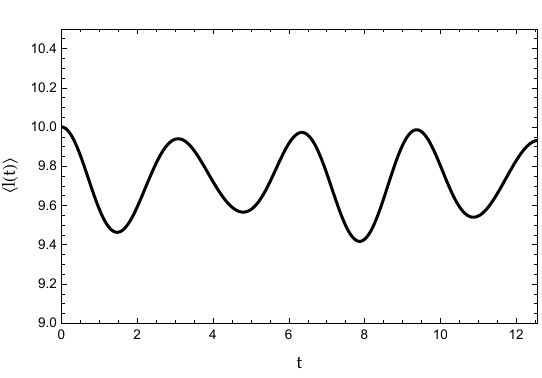}  
\caption{Similar to Fig.~\ref{ztplot0s}, but with $F=1$. }
\label{ztplots}
\end{figure}

\begin{figure}
\includegraphics[width=0.95\textwidth]{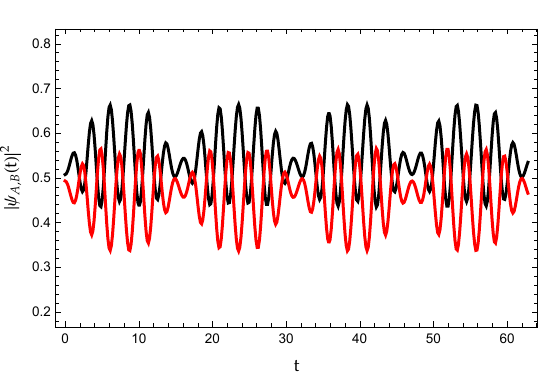} 
\includegraphics[width=0.95\textwidth]{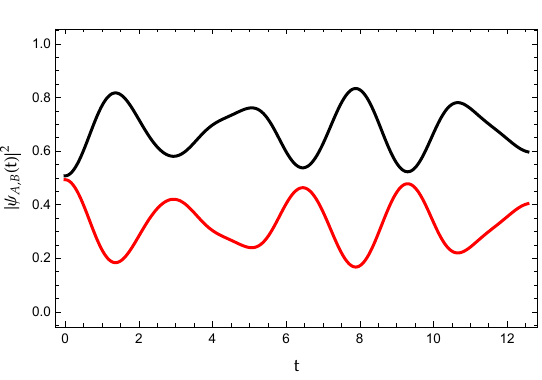}  
\caption{Electron probabilities $ | \Psi_{A} (t)  |^2$ (black)  vs. $ | \Psi_{B} (t)  |^2$ (red)  as defined in Eq.\eqref{PsiAB} for  a $20$-site finite system under the parameters: $\Delta_a=5$, $\Delta_b=1$, $F =  0.2$ (upper panel) and $F=1$ (lower panel). }
\label{probplot}
\end{figure}

\begin{figure}
\includegraphics[width=0.95\textwidth]{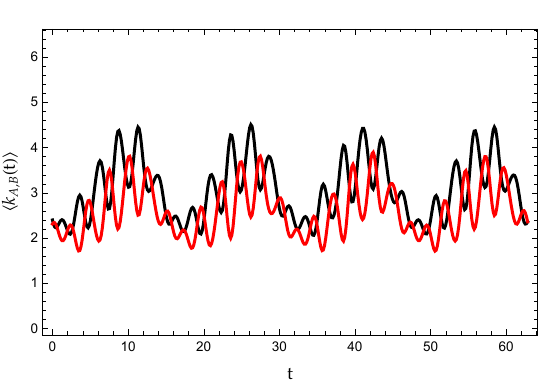} 
\includegraphics[width=0.95\textwidth]{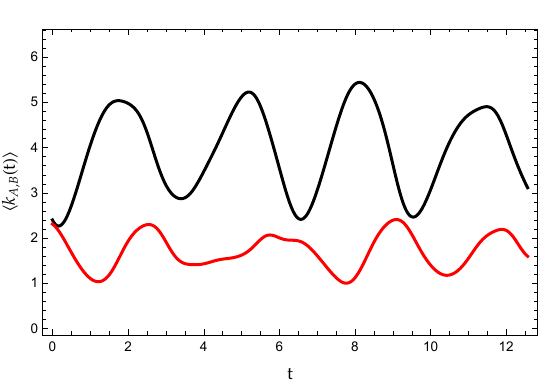}   
\caption{Expectation value of electron momentum  for type-A wave $ \langle  k_{A} (t) \rangle$ (black)  vs. type-B wave $ \langle  k_{B} (t) \rangle$ (red) for a $20$-site finite system with parameters: $\Delta_a=5$, $\Delta_b=1$, $F =  0.2$ (upper panel) and $F=1$ (lower panel).  }
\label{ktplot}
\end{figure}

 In general, the time evolution of single-particle excitation state  in diatomic tight-binding model can be solved numerically by discretizing time,
\begin{equation}
| \Psi (t) \rangle = \lim_{ \substack{ N_t \rightarrow \infty  \\ \delta t \rightarrow 0   } }  e^{- i \hat{H}_{sv} ( N_t \delta t )  \delta t }  \cdots    e^{- i \hat{H}_{sv} (  \delta t )  \delta t }   | \Psi (0) \rangle,
\end{equation}
where $t =  N_t \delta t$ is held fixed at the limit of $N_t \rightarrow \infty$ and  $\delta t \rightarrow 0$.
The matrix representation of $\hat{H}_{sv} (t)$ with periodic boundary condition, $|0 \rangle = | N \rangle $,  is  given by
 \begin{equation}
\hat{H}_{sv}  (t)  = \begin{bmatrix} 
 0 &  - \frac{\Delta_a}{4} & 0 &   \cdots     &   - \frac{\Delta_b}{4}   \\
  - \frac{\Delta_a}{4} & F(t)  &  - \frac{\Delta_b}{4}   & \cdots        &0 \\
 0&  - \frac{\Delta_b}{4} & 2 F (t) & \cdots           &0 \\
\cdots &  \cdots &  \cdots &     \cdots  & \cdots \\
  - \frac{\Delta_b}{4}   & 0& 0&   \cdots  & (N-1)  F(t)   
 \end{bmatrix}  . \label{Hftmatrixrep}
\end{equation}
The eigen-energy solutions for a $20$-site diatomic system with varying $F$ values are plotted in Fig.~\ref{Ekplots} to demonstrate the formation of Wannier-Stark ladders as the field strength $F$ is increased. The expectation value of site-$l$ can be defined on each atom type, 
\begin{equation}
 \langle  l_{A,B} (t) \rangle   = 2 \sum_{l  =(A) even ,  (B) odd }   l \, | \psi(l ,t) |^2 ,
 \label{lab}
\end{equation}
where the factor of $2$ is to take into account the fact that the normalization of each individual type of atom chain is $1/2$ when the tunneling effect is absent. The expectation value of site-$l$ for both types in the tight-binding diatom is simply the average of the two, 
\begin{equation}
 \langle  l (t) \rangle = \frac{ \langle  l_{A} (t) \rangle  +  \langle  l_{B} (t) \rangle }{2}.
 \label{lave}
 \end{equation}
They are plotted in Fig.~\ref{ztplot0s} and Fig.~\ref{ztplots} with a gaussian initial wavefunction,
\begin{equation}
\psi(l, 0) =  (2\pi)^{-\frac{1}{4}} e^{- \frac{(l -\frac{N}{2})^2}{4}} .
\end{equation}
We see that the Bloch oscillation in two-band tight-binding system with unequal hopping exhibits a more complex pattern than that of equal coupling:  not only the type-A and type-B waves oscillate separately moving against each other (almost completely out of phase), but there is a tunneling effect  in the  expectation value of site-$l$ of the entire 1D chain that also oscillates, as indicated in the bottom panel of Fig.~\ref{ztplot0s} and Fig.~\ref{ztplots}.

The collective electron probability for each atom type can be illustrated by summing over individual ones,
\begin{equation}
 | \Psi_{A,B} (t)  |^2= \sum_{l  = (A) even, (B) odd}   | \psi(l ,t) |^2.
 \label{PsiAB}
\end{equation}
An example plot of this probability is shown in Fig.~\ref{probplot}.
We can also define expectation value of individual momentum,
\begin{equation}
\langle k_{A,B} (t) \rangle =  \sum_{k = \frac{2\pi }{N} n }^{n\in[0, N-1]}  \widetilde{\psi}^*_{A,B}(k, t) \, k \,\widetilde{\psi}_{A,B}(k, t) ,
\end{equation}
where $\widetilde{\psi}_{A,B}(k, t) $ are Fourier transform of position-space wavefunctions,
\begin{equation}
\widetilde{\psi}_{A,B}(k, t) = \frac{\sqrt{2}}{\sqrt{N}}  \sum_{ l = (A) even; (B) odd  } e^{- i k l } \psi (l , t).
\end{equation}
The factor $\sqrt{2}$ is to  normalize $\widetilde{\psi}_{A,B}(k, t) $ for each type of atom chain. The two types of atom chains moving against each other are illustrated in the plot of $ \langle  k_{A,B} (t) \rangle$  in Fig.~\ref{ktplot}.

\section{Extension to higher dimensions}
\label{sec:2Dextension}

The statevector representation of the diatomic tight-binding model Hamiltonian can be generalized into higher spatial dimensions in a straightforward manner. Take the 2D single-particle state Hamiltonian as an example. Its statevector basis representation has the typical form,
\begin{equation}
\hat{H} = \sum_{ \mathbf{ l},  \mathbf{ l}' } h_{ \mathbf{ l} ,  \mathbf{ l}' } |  \mathbf{ l} \rangle  \langle   \mathbf{ l}' |,
\end{equation}
where $h_{ \mathbf{ l} ,  \mathbf{ l}' }$ are the coefficients that depend on $( \mathbf{ l} ,  \mathbf{ l}')$,  and the site-$\mathbf{ l}$ in 2D is given by $\mathbf{ l} = (l_x, l_y)$.  In cases where the coefficients $ h_{ \mathbf{ l} ,  \mathbf{ l}' }$ can be factorized into the product of x and y components,
\begin{equation}
h_{ \mathbf{ l} ,  \mathbf{ l}' } = h^{(x)}_{  l_x, l'_x } h^{(y)}_{ l_y, l'_y  },
\end{equation}
the 2D Hamiltonian can be written in terms of tensor product of two terms,
\begin{equation}
\hat{H} =  \hat{H}_x \otimes \hat{H}_y,  
\end{equation}
where 
\begin{equation}
 \hat{H}_x = \sum_{l_x, l'_x}  h^{(x)}_{  l_x, l'_x }  | l_x \rangle  \langle  l'_x |,
\end{equation}
and  $ \hat{H}_y $ is defined in a similar way. Both $ \hat{H}_x$ and  $ \hat{H}_y$ are 1D Hamiltonian that can be mapped into quantum circuits on a set of quantum registers in the same fashion as described in Sec.~\ref{sec:quantumcircuits}. As a specific example, for non-interacting of 1D two-band tight-binding model Hamiltonian   $\hat{H}_{sv} (t)$ in Eq.(\ref{Hftonebody}), the 2D extension of Hamiltonian is thus given by
\begin{equation}
\hat{H}^{(2D)}_{sv} (t) = \hat{H}^{(x)}_{sv} (t) \otimes I^{(y)} +    I^{(x)} \otimes \hat{H}^{(y)}_{sv} (t) ,
\end{equation}
where superscripts $(x,y)$ are used to label dimension of each term. Hence a 2D Hamiltonian can be mapped into two sets of quantum registers, one set is used to describe dynamics in $x$-direction and another set in $y$-direction. The interacting two-particle state Hamiltonian can be handled in a similar way by extending sets of quantum registers.

\end{document}